\documentclass[review]{elsarticle}

\usepackage[utf8]{inputenc}
\usepackage{float}
\usepackage{graphicx}
\usepackage{amsmath}
\usepackage{subfig}
\usepackage{caption}
\usepackage{appendix}
\usepackage{lineno,hyperref}
\usepackage{natbib,doi}
\usepackage{todonotes}
\usepackage[nolist,nohyperlinks]{acronym}

\begin{acronym}
    \acro{ABIDE I}{Autism Brain Imaging Data Exchange I}
    \acro{TD}{typically developing}
    \acro{ASC}{Autism Spectrum Condition}
    \acro{HC}{healthy control}
    \acro{MRI}{magnetic resonance images}
    \acro{SAM}{Statistical Agnostic Mapping}
    \acro{SPM}{Statistical Parametric Mapping}
    \acro{VBM}{voxel-based morphometry}
    \acro{TBV}{total brain volume}
    \acro{GM}{grey matter}
    \acro{WM}{white matter}
    \acro{SBM}{surface-based morphometry}
    \acro{SVM}{support vector machines}
    \acro{CAD}{computer-aided diagnosis}
    \acro{CNN}{Convolutional Neural Network}
    \acro{fMRI}{functional magnetic resonance imaging}
    \acro{sMRI}{structural magnetic resonance imaging}
    \acro{DNN}{Deep Neural Network}
    \acro{GCN}{Graph Convolutional Network}
    \acro{AAL}{Automated Anatomical Labeling}
    \acro{ROI}{regions of interest}
    \acro{CV}{cross-validation}
    \acro{FWE}{Family-wise Error}
    \acro{PLS}{Partial Least Squares Regression}
    \acro{GLM}{General Linear Model}
    \acro{ReML}{restricted maximun likelihood}
    \acro{FES}{Feature Extraction and Selection}
    \acro{ABIDE}{Autism Brain Imaging Data Exchange}
\end{acronym}

\begin{document}

\begin{frontmatter}

\title{Unraveling the Autism spectrum heterogeneity: Insights from ABIDE I Database using data/model-driven permutation testing approaches}

\author[mymainaddress]{F.J. Alcaide}
\author[mymainaddress]{I.A. Illan}
\author[mymainaddress]{J.  Ram{\'i}rez}
\author[mymainaddress,mysecondaryaddress,mythirdaddress]{J.M. Gorriz\corref{mycorrespondingauthor}}
\ead{gorriz@ugr.es}
\cortext[mycorrespondingauthor]{Corresponding author}

\address[mymainaddress]{DaSCI Institute, University of Granada, Spain}
\address[mysecondaryaddress]{Dpt. of Psychiatry, University of Cambridge, UK}
\address[mythirdaddress]{ibs.Granada, Granada, Spain}


\begin{abstract}
\ac{ASC} is a neurodevelopmental condition characterized by impairments in communication, social interaction and restricted or repetitive behaviors. Extensive research has been conducted to identify distinctions between individuals with ASC and neurotypical individuals. However, limited attention has been given to comprehensively evaluating how variations in image acquisition protocols across different centers influence these observed differences. This analysis focuses on \ac{sMRI} data from the Autism Brain Imaging Data Exchange I (ABIDE I) database, evaluating subjects' condition and individual centers to identify disparities between ASC and control groups. Statistical analysis, employing permutation tests, utilizes two distinct statistical mapping methods: Statistical Agnostic Mapping (SAM) and Statistical Parametric Mapping (SPM). Results reveal the absence of statistically significant differences in any brain region, attributed to factors such as limited sample sizes within certain centers, nuisance effects and the problem of multicentrism in a heterogeneous condition such as autism. This study indicates limitations in using the ABIDE I database to detect structural differences in the brain between neurotypical individuals and those diagnosed with \ac{ASC}. Furthermore, results from the SAM mapping method show greater consistency with existing literature.

\end{abstract}


\end{frontmatter}

\section{Introduction}
\ac{ASC}, also referred to as autism, is a neurodevelopmental disorder with significant implications for communication, social interaction, and behavioral patterns \cite{lombardo_big_2019, masi_overview_2017, szatmari_heterogeneity_1999}. The disorder exhibits a complex nature, encompassing a broad spectrum of symptoms and severity levels, thereby posing challenges in terms of diagnosis and treatment \cite{hodges_autism_2020}. In the previous edition of the American Psychiatric Association's Diagnostic and Statistical Manual of Mental Disorders (DSM-IV) \cite{bell_dsm-iv_1994}, autism was categorized into five distinct subtypes: Autistic Disorder, Rett Syndrome, Asperger Syndrome, Childhood Disintegrative Disorder, and Pervasive Developmental Disorder Not Otherwise Specified. However, the fifth edition of the DSM (DSM-5) introduced updated recommendations for the classification of \ac{ASC}. Presently, the diagnostic range for \ac{ASC} is unified under a single spectrum, eliminating the subtypes. The DSM-5 \cite{american_psychiatric_association_guiconsulta_2014} incorporates a generalized \ac{ASC} category, omitting Rett Syndrome and Childhood Disintegrative Disorder. This revised classification framework, centered on a unified spectrum, aims to improve the comprehensive representation of the diverse manifestations observed in \ac{ASC}. It also facilitates the diagnosis of cases characterized by subtle yet intrinsic features of the disorder. Consequently, individuals with \ac{ASC}, including those with milder phenotypes, can gain access to tailored therapies aligned with their specific requirements.

The long-term consequences of \ac{ASC} exhibit increased severity, contributing to a higher susceptibility of affected individuals to medical complications \cite{jones_description_2016}. Additionally, \ac{ASC} demonstrates a significant gender imbalance, with male-to-female ratios of 1.8:1 in adults and 3.5:1 in children and adolescents \cite{rutherford_gender_2016}. Nonetheless, our understanding of the inherent cerebral distinctions between neurotypical individuals and those with \ac{ASC} remains limited. Efforts have been made over the years to address this knowledge gap \cite{ecker_neuroanatomy_2017}, prompting researchers to employ neuroimaging methodologies for investigating the structural and functional aspects of the brain in individuals with \ac{ASC} \cite{verhoeven_neuroimaging_2010}. 

\subsection{Related Works}
Within the existing literature, a considerable number of research papers have focused on investigating the anatomical differences in the brains of individuals with autism compared to neurotypical individuals\cite{rane_connectivity_2015,del_casale_neuroanatomical_2022, rafiee_brain_2022}. These studies aim to identify specific brain regions that exhibit significant differences, striving to achieve a high level of accuracy in identifying and understanding the neurological characteristics associated with autism \cite{ali_role_2022}.

Studies such as the one conducted by Segovia et al. \cite{segovia_identifying_2014}, the focus was on identifying differences among three distinct groups: control participants, individuals with \ac{ASC}, and unaffected siblings of individuals with \ac{ASC}. They employed a univariate methodology using \ac{MRI}, combining a reflector approach and a \ac{SVM} classifier to detect differences between the groups.

McAlonan et al. \cite{mcalonan_mapping_2005} conducted a comprehensive voxel-based volumetric analysis of the entire brain using automated techniques. They compared children with \ac{ASC} to \ac{TD} children. The findings of the study indicated abnormalities in the anatomy and connectivity of the limbic-striatal brain systems. These observed abnormalities may contribute to differences in brain metabolism and contribute to the behavioral characteristics associated with autism.

Gori et al. \cite{gori_gray_2015} conducted sMRI analysis of children with \ac{ASC} and \ac{HC} using \ac{SPM} and Freesurfer to extract morphometric brain information, and a \ac{SVM} classifier to localize the brain regions contributing to classification.

Li et al. \cite{li_reduction_2019} analyzed \ac{GM} volume in children with \ac{ASC} compared to \ac{HC}, particularly focusing on children with Low Functioning Autism (LFA) \cite{lincoln_assessment_1995}, obtaining findings regarding a reduction in GM volume in several brain regions.

\section{Machine learning methods in ASC}

Classical machine learning (ML) algorithms have extensively been used to extract distinctive features from neuroimages in order to discern variations between subjects\cite{ECKER201044}. Several studies have utilized \ac{VBM} \cite{ashburner_voxel-based_2000} to examine and compare \ac{TBV} and regional disparities in individuals with \ac{ASC} \cite{deramus_anatomical_2014}. In  \cite{riddle_brain_2017}, it was observed that \ac{TBV} and gray matter volumes increased by 1-2\% in individuals with \ac{ASC}. However, this finding only attained statistical significance when the entire dataset was considered, without matching the images by age and sex. A common finding in \cite{piven_regional_1996, piven_magnetic_1992, piven_mri_1995} is the identification of a significantly larger median sagittal brain area in individuals with \ac{ASC} compared to control subjects. Furthermore, these studies indicated that the increase in brain volume \cite{lainhart_macrocephaly_1997} is not a uniform phenomenon but varies across the frontal, temporal, parietal, and occipital lobes. In contrast, \cite{aylward_effects_2002} reported no discernible differences in brain volume among subjects older than 12 years. However, an increase in head circumference was observed in individuals with \ac{ASC}, leading to the conclusion that it stemmed from an expansion in brain volume during childhood.

In \cite{gorriz_machine_2019},  a partial least squares feature extractor was employed to demonstrate disparities in regional brain structure between genders and neurological conditions. The study highlighted the influence of gender on \ac{ASC}, even when utilizing a balanced dataset for classification. Similarly, \cite{zhang_revisiting_2018} investigated age and gender dependencies using volumes of \ac{GM}, \ac{WM}, and subcortical structures, yielding positive outcomes.

Other diverse approaches encompassed \ac{SBM} \cite{jiao_predictive_2010}, analysis of words and phrases from children's assessments \cite{maenner_development_2016}, and the utilization of multiple \ac{SVM} \cite{boser_training_1992, mammone_support_2009} to enhance the final classification \cite{bi_classification_2018}. In \cite{ali_role_2022}, a \ac{CAD} system was developed, involving the extraction of the cerebral cortex from \ac{sMRI} and the creation of personalized neuroatlases that describe specific developmental patterns of the autistic brain.

\subsection{Novel Advances in neuroimaging based on ML}
Similar to classical analysis approaches, neural networks have been utilized to leverage structural and functional connectivity patterns for the classification of individuals with \ac{ASC}. The \ac{ABIDE I} dataset was employed as a common database in these studies. In \cite{heinsfeld_identification_2017}, subjects were classified based on their brain activation patterns, achieving a 70\% accuracy in identifying patients with \ac{ASC} compared to controls. In \cite{sherkatghanad_automated_2020}, the focus was on automatic detection of \ac{ASC} using \ac{CNN}) with a brain imaging dataset consisting of resting-state \ac{fMRI} data. The model successfully classified subjects based on patterns of functional connectivity, achieving an accuracy of 70.22\% in detecting \ac{ASC}. \cite{yang_deep_2020} implemented a design incorporating four distinct hidden layer configurations of \ac{DNN} models to classify \ac{ASC} from \ac{TD}) individuals. The model achieved an accuracy of 75.27\%, recall of 74\%, and precision of 78.37\%. In \cite{kong_classification_2019}, individual brain networks were constructed for each subject, and a \ac{DNN} classifier was employed with the input consisting of the 3000 most relevant features. The results demonstrated an accuracy of 90.39\% and an area under the receiver operating characteristic curve (AUC) of 0.9738 for \ac{ASC}/\ac{TD} classification. Ms et al. \cite{ms_darkasdnet_2021}introduced DarkASDNet, a new approach utilizing \ac{fMRI} data for predicting the binary classification between \ac{ASC} and \ac{TD} in the ABIDE-I, NYU dataset. The proposed method achieved state-of-the-art accuracy of 94.70\% in the classification task. In \cite{selcuk_nogay_diagnostic_2023}, a grid search optimization (GSO) algorithm was applied to optimize hyperparameters in Deep Convolutional Neural Networks (DCNN) used in the system. As a result, the proposed diagnostic method based on \ac{sMRI} achieved an exceptional success rate of 100\% in identifying \ac{ASC}. 

In \cite{khosla_ensemble_2019}, the authors investigated the impact of stochastic parcellations of the brain by employing an ensemble architecture in conjunction with a \ac{CNN}. This approach facilitated the combination of predictions from different parcellations. Temporal and spatial information from \ac{fMRI} images were leveraged in \cite{zhang_survey_2020}. A sliding window technique was applied to create two-channel images, which served as the input to a \ac{CNN}. This enabled the exploration of dynamic patterns in the data. Zhao et al. \cite{zhao_3d_2018} introduced an effective 3D \ac{CNN} framework that derived discriminative and significant spatial brain network overlap patterns. This framework proved valuable for distinguishing between individuals with \ac{ASC} and \ac{TD}.

In recent advancements of deep learning for neuroimaging, particular attention has been given to \ac{GCN}. These networks enable classification not only at the individual level but also consider the relationships within the entire population of data. In \cite{arya_fusing_2020, rakic_improving_2020}, the combination of functional and structural \ac{MRI} information was employed to enhance the classification performance. Accuracies of 64.23\% and 85.06\% were achieved, respectively. Specifically,  \cite{arya_fusing_2020} utilized the fusion of information to construct the edges and nodes of the \ac{GCN}. Furthermore, in \cite{anirudh_bootstrapping_2018}, an ensemble of weakly trained \ac{GCN}s was utilized to enhance performance robustness against architectural changes. This approach aimed to improve the overall stability and reliability of the classification results. Through the implementation of this method, an accuracy of 70.86\% was achieved.

This paper aims to investigate the structural MRI data within the multicenter and international ABIDE I database \cite{di_martino_autism_2014}, focusing on its relevance to characterizing autism. It examines patients in a comprehensive manner based on their condition, as well as investigates each center individually. The analysis will employ rigorous statistical methodologies, including a permutations test, and two established neuroimaging mapping techniques, namely \ac{SAM} \cite{gorriz_statistical_2021} and \ac{SPM} \cite{friston_statistical_1994, penny_statistical_2011}. The primary objective is to identify and validate any brain regions demonstrating significant differences between individuals diagnosed with autism and healthy controls. Furthermore, the study underscores the inherent challenges posed by highly heterogeneous data within multicenter databases like ABIDE, particularly in the context of analyzing a disorder as diverse as autism\cite{mottron_autism_2020}.
\section{Materials and Methods}

\subsection{ABIDE I dataset}

Comprehensive and extensive datasets are required to further our understanding of the underlying brain mechanisms associated with autism and the intricate and heterogeneous nature of the disorder. To address this need, the \ac{ABIDE I} initiative was established. \ac{ABIDE I} aims to collect functional and structural brain imaging data from multiple research laboratories worldwide, providing a valuable resource for advancing our knowledge of \ac{ASC}.

The \ac{ABIDE I} database comprises magnetic resonance images shared by 20 international centers. For our analysis, we accessed a dataset consisting of 1032 images, with 527 images corresponding to \ac{HC} and 505 images representing individuals with \ac{ASC}, as summarized in Table \ref{ABIDE}. The brain images of each subject sourced from the \ac{ABIDE I} database underwent preprocessing and segmentation procedures, with only the complete \ac{GM} map of each image being utilized. Subsequently, the brain volumes were partitioned into 116 \ac{ROI} using a brain atlas established by the \ac{AAL} method \cite{tzourio-mazoyer_automated_2002}. Within the article \cite{sun_mining_2009}, there is a table (Table 2) which provides the names and corresponding numbers, according to the predefined \ac{AAL} order, of the 116 \ac{ROI} that are employed in the study.

\begin{table}
    \centering
    \caption{Number of brain images contributed by each center, categorized according to their condition, specifically control (HC) or autism (ASC).}
    \vspace{0.3cm}
    \begin{tabular}{l c c}
    \hline
    Sites & HC & ASC \\
    \hline
    CALTECH & 18 [14M/4F] & 19 [15M/4F] \\
    CMU & 13 [10M/3F] & 14 [11M/3F] \\
    KKI & 28 [20M/8F] & 20 [16M/4F] \\
    LEUVEN\_1 & 15 [15M/0F] & 14 [14M/0F] \\
    LEUVEN\_2 & 19 [14M/5F] & 15 [12M/3F] \\
    MAX\_MUN & 28 [27M/1F] & 24 [21M/3F] \\
    NYU & 100 [74M/26F] & 75 [65M/10F] \\
    OHSU & 15 [15M/0F] & 13 [13M/0F] \\
    OLIN & 14 [14M/0F] & 20 [17M/3F] \\
    PITT & 26 [22M/4F] & 30 [26M/4F] \\
    SBL & 15 [15M/0F] & 15 [15M/0F] \\
    SDSU & 22 [16M/6F] & 14 [13M/1F] \\
    STANFORD & 17 [13M/4F] & 18 [14M/4F] \\
    TRINITY & 25 [25M/0F] & 22 [22M/0F] \\
    UCLA\_1 & 30 [27M/3F] & 41 [35M/6F] \\
    UCLA\_2 & 13 [11M/2F] & 13 [13M/0F] \\
    UM\_1 & 54 [37M/17F] & 53 [45M/8F] \\
    UM\_2 & 21 [20M/1F] & 13 [12M/1F] \\
    USM & 26 [26M/0F] & 44 [44M/0F] \\
    YALE & 28 [20M/8F] & 28 [20M/8F] \\
    \hline
    \end{tabular}
    \label{ABIDE}
\end{table}

\subsection{Preprocessing and outliers detection}
The \ac{MRI} images obtained from the \ac{ABIDE I} database were subjected to processing and segmentation using the \ac{SPM} software. While \ac{SPM} was primarily developed for functional imaging, it also offers functionalities for spatial realignment, smoothing, and normalization in the standard T1-weighted image space. The entire process was conducted utilizing the \ac{VBM} Toolbox within \ac{SPM} \cite{vbm8_manual}. The preprocessing steps, including co-registration and segmentation, as well as the specific parameters employed at each stage, can be summarized as follows:

\begin{enumerate}
    \item Preprocessing
    \begin{itemize}
        \item The process used tissue probability maps to guide the analysis. The International Consortium for Brain Mapping (ICBM) provide these maps, which are derived from 452 T1-weighted scans that were aligned with an atlas space, corrected for scan inhomogeneities, and classified into three tissue types: \ac{GM}, \ac{WM}, and Cerebrospinal Fluid (CSF). The resulting maps were used to estimate a nonlinear deformation field that best aligns the tissue probability maps with each individual subject's image. The images were resized to 79 x 95 x 79 voxels with voxel sizes of 2 mm (sagital) x 2 mm (coronal) x 2 mm (axial).
        \item A mutual information affine registration with the tissue probability maps was used to achieve approximate alignment.
        \item Spatial normalization was based on a high-dimensional Dartel normalization and used standard Dartel template provided by VBM 8
    \end{itemize}
    \item Segmentation
    \begin{itemize}
        \item For each tissue class (GM, WM, CSF), the intensity distribution was represented using a certain number of Gaussians, 2 in this case. The use of multiple components per tissue allows to reckon partial volume effects and deep GM differing from cortical GM.
        \item A very light bias regularization was performed to correct smooth, spatially varying artifacts that modulates the intensity of the images.
        \item A spatial adaptive non local means denoising filter is applied to the data in order to remove noise while preserving edges. The smoothing filter size is automatically estimated based on the local variance in the image.
        \item Skull stripping was performed by using SPM-VBM tool and VBM templates.
    \end{itemize}
\end{enumerate}

As a result of the preprocessing and segmentation procedures, probability maps were generated for each MRI image in the database. These maps assigned values within the range of 0 to 1 to each voxel, representing the probability of that voxel belonging to specific tissues, such as \ac{WM}, \ac{GM}, or CSF. However, for the purpose of our study, only the \ac{GM} map derived from the images will be utilized.

\subsection{Statistical Parametric Mapping}

\begin{figure} [H]
    \centering
    \includegraphics[scale=0.41]{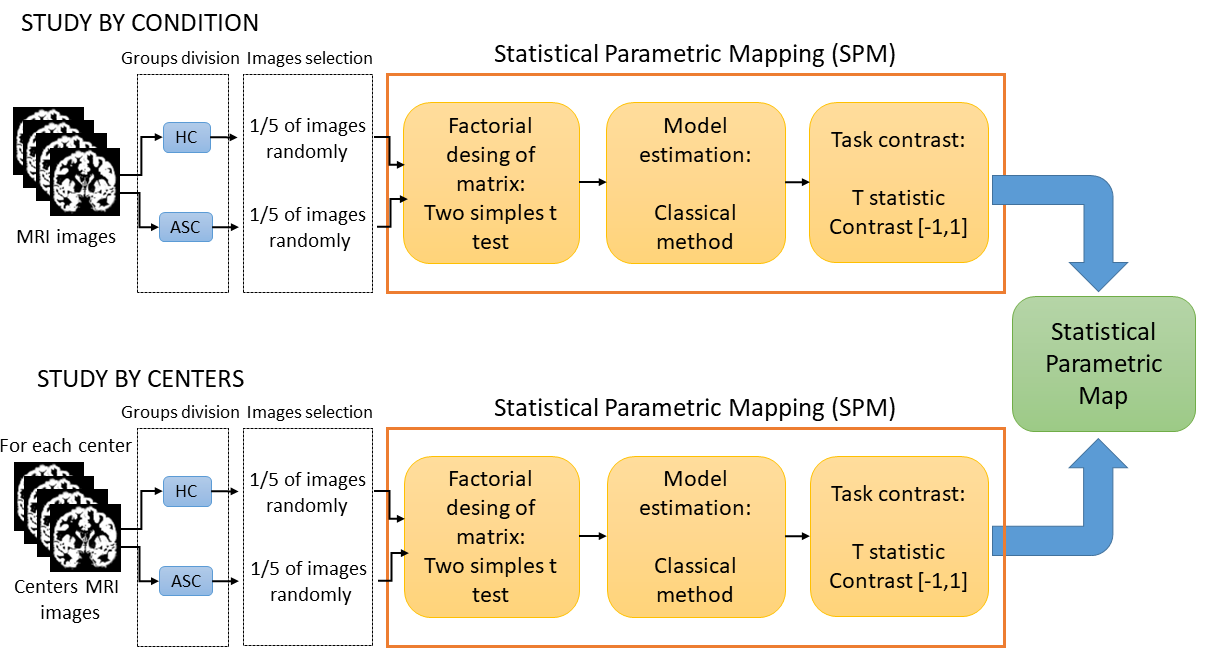}
    \caption{The block diagram illustrates the \ac{MRI} analysis process using \ac{SPM} and its three main steps. The diagram specifically focuses on the comparison between \ac{HC} and individuals with \ac{ASC}. For the comparison between healthy controls themselves (\ac{HC} vs \ac{HC}), the group division involves creating two separate \ac{HC} groups while maintaining the remaining stages of the analysis process.}
    \label{diagrama_spm}
\end{figure}

\ac{SPM}, or Statistical Prametric Mapping \cite{friston_statistical_1994, penny_statistical_2011}, is a neuroimaging tool that utilizes voxel-wise statistical analysis to examine and compare structural and functional brain data in relation to different conditions or tasks. The statistical analysis of \ac{MRI} data employs a univariate mass approach based on \ac{GLM}. This approach entails the specification of a \ac{GLM} design matrix, which in our case involves utilizing a two-sample $t$-test to divide the \ac{MRI} images into two equally-sized groups. Subsequently, the model is estimated through the classical method, employing \ac{ReML} to estimate the parameters. This estimation assumes that the error correlation structure is uniform across all voxels. Following the estimation, specific parameter profiles are subjected to testing using T statistics with contrasts assigned values of +1 and -1 for each respective image class. To mitigate the risk of false positive (FP) results in the context of multiple comparisons, a whole-brain cluster-level \ac{FWE} correction was applied, maintaining a significance level ($\alpha$) of 0.05, rejecting the null hypothesis if $p <$ 0.05. This correction technique helps ensure that only statistically significant findings are considered. Figure \ref{diagrama_spm} presents a block diagram illustrating the analysis process with \ac{SPM}, highlighting the main steps involved.

\subsection{Statistical Agnostic Mapping}
\ac{SAM}, or Statistical Agnostic Mapping \cite{gorriz_statistical_2021}, is a non-parametric ML method used for evaluating neuroimaging data at either the voxel or regional level. It addresses the issue of unstable risk estimates that arise from limited sample sizes by employing \ac{CV} techniques in ML. Moreover, \ac{SAM} offers an alternative approach for mapping $p-$values that are corrected for \ac{FWE} under \emph{the worst case}, ensuring the reliability of inferential statistics for hypothesis testing. The methodology of \ac{SAM} is grounded in the data and employs concentration inequalities to test opposing hypotheses or compare different models. \ac{SAM} has been shown to be a highly competitive and complementary method to the \ac{SPM} framework, which is widely accepted by the neuroimaging community. \ac{SAM} generates activation maps similar to those obtained through voxel-wise analysis in \ac{SPM}, but with a focus on regions of interest. It has been extensively developed and tested in scenarios characterized by a small sample-to-dimension ratio and varying effect sizes, including large, small, and trivial effects.

\subsubsection{Feature Extraction and Selection}
The \ac{SAM} procedure can be summarized into several steps as follows: 

\begin{enumerate}
    \item The data will be prepared by designing a comparison between groups, referred to as the design matrix. \ac{ROI} will be selected for analysis, focusing on specific areas within the subjects' brains.
    
    \item For each \ac{ROI}, a feature extraction and selection stage will be conducted to derive the feature space. This involves identifying relevant features that contribute to the classification task. A \ac{SVM} will be trained using Empirical Risk Minimization (ERM) to estimate the replacement error.

    \item The empirical error or precision will be calculated based on the trained \ac{SVM} model. Additionally, the true precision will be determined in the worst-case scenario with a probability of 1 - $\delta$. This step helps assess the accuracy and reliability of the results.

    \item The $z-$test statistic will be computed for each true precision to evaluate their significance. This statistical test assesses whether the observed results deviate significantly from what would be expected by chance, providing insights into the meaningfulness of the findings.
\end{enumerate}

\begin{figure} [H]
    \centering
    \includegraphics[scale=0.41]{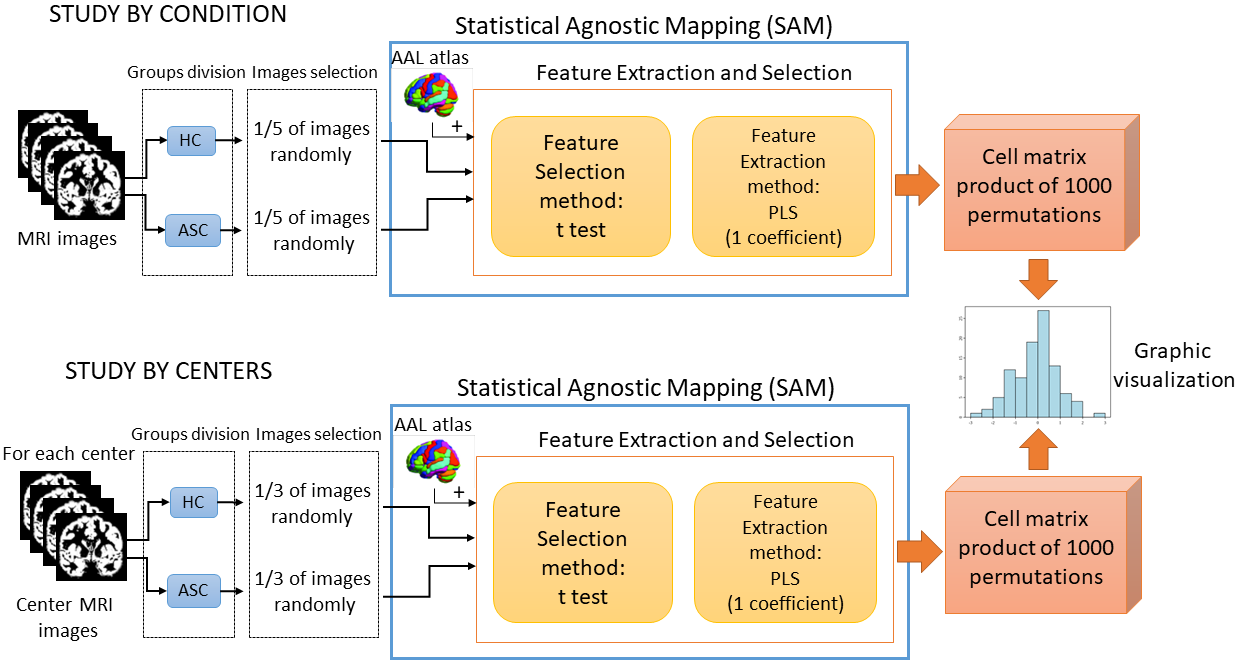}
    \caption{The block diagram illustrates the \ac{MRI} analysis process incorporating \ac{SAM} and the feature extraction and selection stage. Specifically, the diagram represents the comparison between \ac{HC} and individuals with \ac{ASC}. As for the comparison between \ac{HC} themselves (\ac{HC} vs \ac{HC}), the group division will involve creating two separate \ac{HC} groups, while retaining the remaining stages of the analysis process.}
    \label{diagrama_sam}
\end{figure}

These steps collectively form the \ac{SAM} procedure, allowing for the identification and evaluation of significant differences between groups in the selected regions of interest. In this study, default values are used for the SAM implementation (v1.0 \cite{SAMv1}).  The input data for \ac{SAM} will consist of images divided into two classes, namely HC class images vs. ASC class images; and HC class images vs. HC class images. To select relevant features, we will employ a $t-$test. Within each \ac{ROI}, \ac{SAM} will calculate the $t-$statistic value for each voxel, and we will select the 50 voxels (features) with the highest $t-$values. This threshold of 50 voxels has been determined based on the consideration of images with a voxel size of 2x2x2 mm and an atlas comprising 116 regions, as it has consistently yielded favorable outcomes in previous studies\cite{gorriz_statistical_2021}. \ac{PLS} will be used for feature extraction\cite{krishnan_partial_2011}, with a single \ac{PLS} dimension being employed (see \ref{pls_dimension}). \ac{PLS} has demonstrated its effectiveness in capturing relationships between brain activity and experimental design or behavioral measures within a multivariate framework. The complete analysis pipeline, which incorporates \ac{FES}, is illustrated in Figure \ref{diagrama_sam} as a block diagram.

\subsection{Statistical Analysis by permutation testing}
In this article, we employ a two-sample statistical analysis using permutation tests \cite{pesarin_permutation_2010, good_permutation_2000} to identify brain regions that exhibit statistically significant differences. The $p-$value will be calculated using equation (\ref{p value}), where the numerator represents the number of instances where the permutation test statistic $\mathcal{T}$ exceeds the true precision, divided by the total number of permutations conducted. Adding 1 to both the numerator and denominator ensures the inclusion of a fixed statistical value, thereby ensuring the validity of the test.

\begin{equation}
\label{p value}
    p\_value = \frac{[\#\mathcal{T}_\pi \geq \mathcal{T}]+1}{M+1}
\end{equation}
where $M$ is the number of permutations.
\vspace{2mm}

The study involves analyzing brain images using two different approaches: by center and by condition. In each case, two types of comparisons will be conducted: between individuals with autism and controls (\ac{HC} vs \ac{ASC}), and between controls themselves (\ac{HC} vs \ac{HC}). When analyzing by center, these comparisons will be performed individually for each center. The purpose of conducting the study by center and by condition is to minimize the possibility of obtaining FP results caused by external factors unrelated to the disorder, such as variations in the instruments used to obtain the images, data acquisition processes, or other confounding factors. The comparisons of primary importance are those between \ac{HC} and \ac{ASC}, as they involve distinct classes of individuals and determine the presence of significant differences between the groups. However, the comparison of \ac{HC} vs \ac{HC} is included to identify any regions that may exhibit significant differences unrelated to the disorder, possibly due to chance or other external factors. If such regions emerge in both the \ac{HC} vs \ac{HC} and \ac{HC} vs \ac{ASC} comparisons, they can be disregarded as unrelated to the disorder.

To conduct the aforementioned comparisons, a permutation test will be employed, with $M=1000$ permutations performed for each case\cite{OJA2009}. The permutation test is a statistical significance test used to examine differences between groups. It involves calculating the statistic value for all possible rearrangements of observations within the different groups. In each permutation, the brain images will be divided into two groups of equal sizes: \ac{HC} and individuals with \ac{ASC}, or two \ac{HC} groups.

\subsection{Sample size and power calculations}
\begin{figure} 
    \centering
    \includegraphics[scale=0.4]{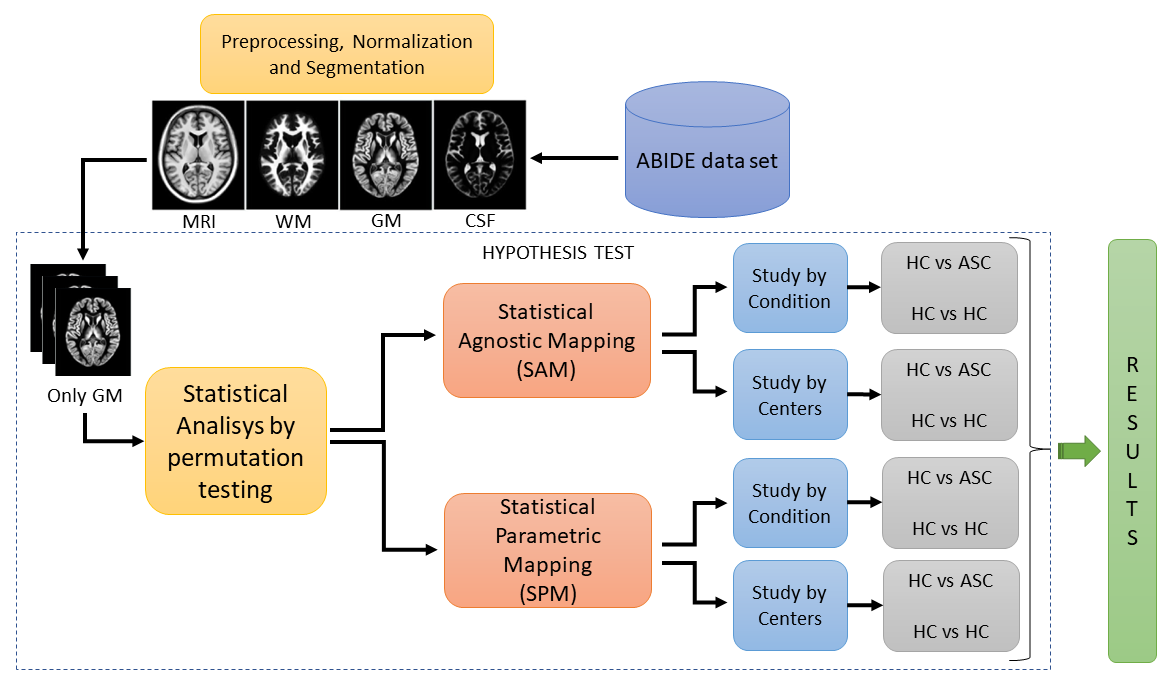}
    \caption{Block diagram about structure of the Autism Brain Imaging Data Exchange I (ABIDE I) dataset analysis.}
    \label{diagrama_completo}
\end{figure}

A specific number of images will be randomly selected from each group for the analysis, assuming  the sample size needed to detect the minimum detectable effect. In the case of the study based on the condition of the patients (\ac{HC} vs \ac{ASC}), approximately $1/5$ of the total number of images contributed by each group will be taken, resulting in 100 images from each group. Conversely, for the study conducted by centers, 1/3 of the images from each group will be selected individually at each center. This ensures a balance between the sample size and the available dataset, creating distinct subgroups in each permutation. The varying sample sizes are a result of the total number of samples available in each case, with the study by centers having a smaller number of available images. In each permutation, the selected images will be compared to one another using the respective mapping method being employed. This involves region-to-region comparisons using \ac{SAM} and voxel-level comparisons using \ac{SPM}. Figure \ref{diagrama_completo} illustrates the framework of the analysis process.

\section{Experimental results}

The initial analysis of the data will be conducted using the data-driven \ac{SAM} method. The objective is to take advantage of the excellent control over FPs provided by this method, in contrast to the optimistic clusterwise inference of parametric methods \cite{eklund_cluster_2016}. 

\subsection{SAM analysis}

Firstly, the analysis will be carried out based on the condition, where all the images contributed by each center will be grouped together, distinguishing between \ac{ASC} and \ac{HC}. With 1000 permutations and a probability of at least $1-\alpha$ set at 0.95, we would expect around 50 FPs by chance under the null hypothesis. Therefore, regions that exhibit a number of significant differences exceeding 50 will be deemed significant (TP), as they fall within the predetermined 95\% confidence level (1 - $\alpha$) in HC vs. ASC comparison. In the results obtained it can discern the presence of numerous regions displaying a notable number of significant differences. Surprisingly, even in the \ac{HC} vs \ac{HC} comparison, where any significant difference should be considered as a FP, regions exhibiting substantial differences are observed. Notably, both comparisons reveal the same significant regions with highly similar values, indicating that many of these differences are not solely attributable to the patient's condition (if any) but rather to other factors. Consequently, in order to investigate whether any of these differences may be attributed to the distinct imaging centers contributing the data, a subsequent study will be conducted focusing on the centers.

\begin{figure}
    \centering
    \subfloat[HC vs ASC permutations comparison of each center with SAM]{
        \label{HC vs ASC permutations comparison}
        \includegraphics[width=\textwidth]{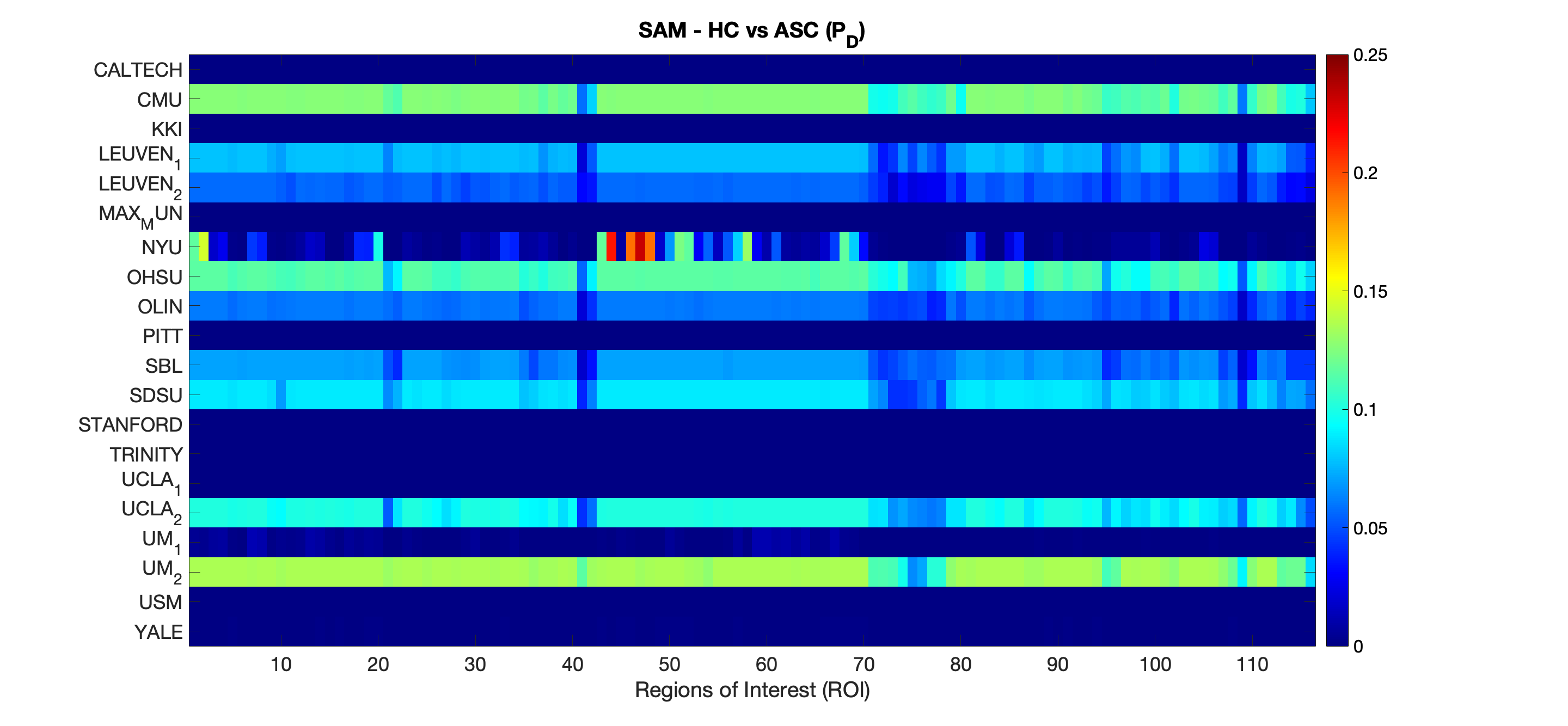}}
        
    \subfloat[HC vs HC permutations comparison of each center with SAM]{
        \label{HC vs HC permutations comparison}
        \includegraphics[width=\textwidth]{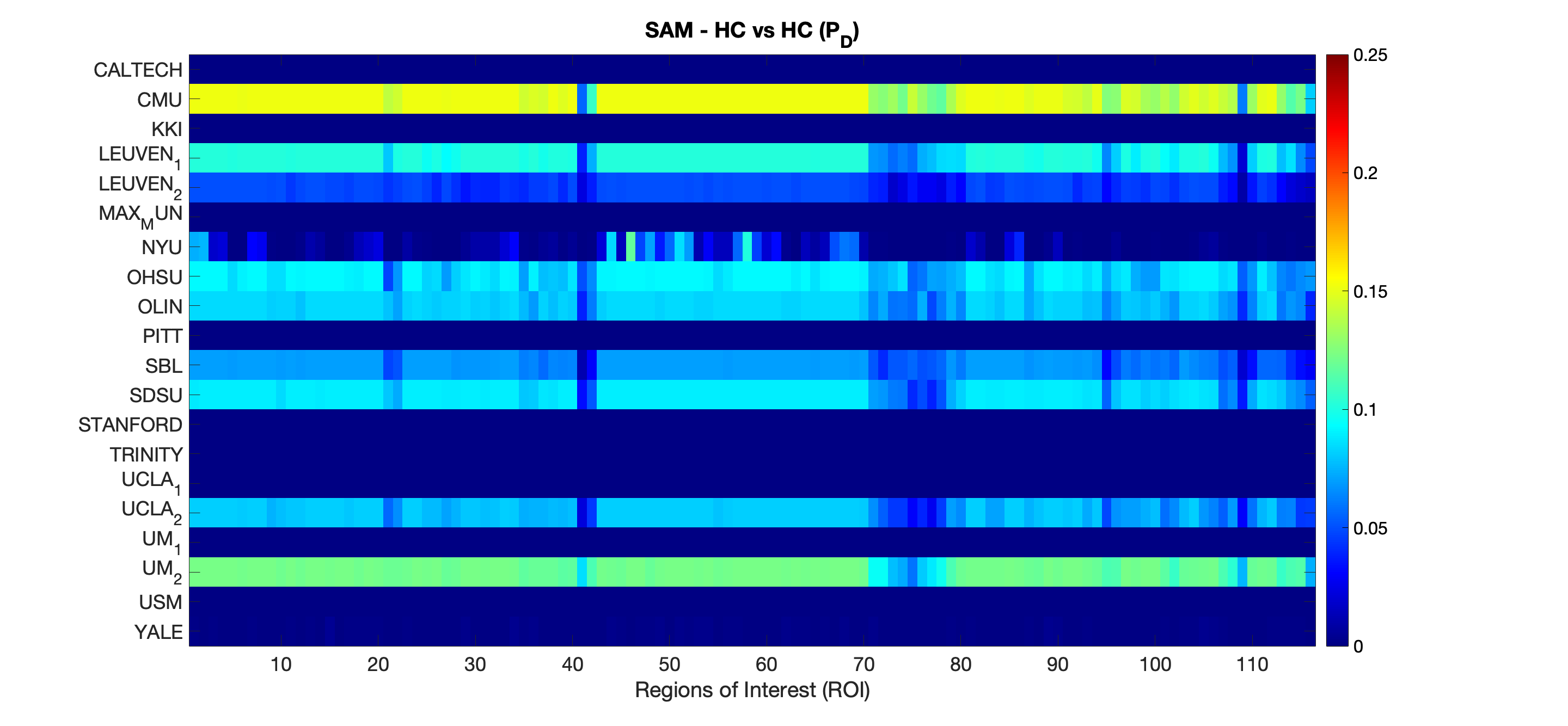}}
    \caption{Estimated probability of detection through permutation testing using the \ac{SAM} mapping method. Each colormap rectangle represents one of the 116 brain regions defined in Table 2 of \cite{sun_mining_2009}. The color map depicts the probability of detection of significant differences observed in the regions during the permutation test.}
    \label{comparison centers}
\end{figure}

The permutation test, along with SAM, is performed individually for each center in both comparisons (see figure \ref{comparison centers} where we display ``probability of detection''). Analyzing the results shown in the aforementioned figure, it is possible to distinguish three different sets of centers that show similar behaviour: (i) Firstly, there are centers where no significant differences are observed in any region, regardless of the comparison being made. While this outcome may be anticipated, as it suggests that no specific brain region distinguishes individuals with \ac{ASC} from \ac{HC}, it could be surprising that none of the regions exhibit even minimal differences across multiple brains, even if these differences are attributable to random factors. (ii) Secondly, there are centers where the \ac{HC} vs \ac{HC} comparison yields no significant differences in any region, similar to the previous case. However, in the \ac{HC} vs \ac{ASC} comparison, almost all regions demonstrated significant differences, with remarkably similar frequencies. This is an exceptionally unusual scenario, as it is highly unlikely for all brain regions to exhibit such consistent and significant differences. (iii) Lastly, the centers NYU, UM\_1, and YALE exhibit more expected results. In both comparisons, these centers show a mixture of regions with significant differences and regions without significant differences. Overall, these findings highlight the variability in results across centers, with some centers showing no significant differences (conservative behavior of the test), others displaying abnormal patterns of significant differences (nuisance effects), and a few demonstrating more ``typical outcomes'' with a mix of significant and non-significant regions in \emph{both comparisons}.

\begin{figure} 
    \centering
    \includegraphics[scale=0.7]{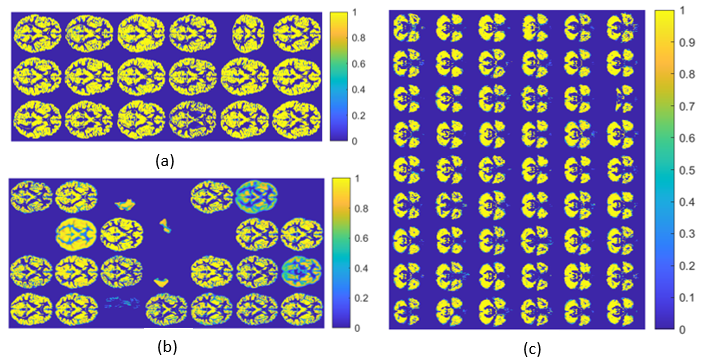}
    \caption{Brain mosaics from different centers to assess the presence of defective images. Specifically, the mosaics of three centers with distinct issues in their images are shown:(a) CALTECH's cerebral mosaic, (b) YALE's cerebral mosaic and (c) UM\_1's cerebral mosaic.}
    \label{mosaico}
\end{figure}

The initial analysis was performed without accounting for the presence of poor quality images, as shown in Figure \ref{mosaico}. Their presence can be attributed to errors in the acquisition or pre-processing pipeline, and affects every group randomly. Therefore, the balance in age or center is not affected by excluding them. The impact of incorporating defective images in the region detection process can be assessed by subsequently removing them from the dataset, thereby assessing the overall influence of noisy data.

\subsubsection{An analysis removing outliers}

By visual inspection, 21 images with defects that made them unusable were detected and discarded.   After conducting subsequent analyses excluding the aforementioned images, it was found that the new results were highly consistent with the previous findings. However, there were notable changes in certain centers. The centers belonging to the second case described earlier now exhibited results where all regions displayed significant differences in both the \ac{HC} vs \ac{ASC} and \ac{HC} vs \ac{HC} comparisons. Additionally, the NYU center showed variations in the significance levels across different regions. On the other hand, the UM\_1 center fell into the second case category, showing no effect results in the \ac{HC} vs \ac{HC} comparison and only minor differences in the \ac{HC} vs \ac{ASC} comparison, which were not significant as they fell below the threshold of 0.05. Similar to UM\_1, the YALE center could also be classified as the second case, as it exhibited various regions with a similar number of differences in both comparisons. However, like UM\_1, none of the regions in the YALE center demonstrated significant differences that surpassed the threshold for significance.

Figure \ref{comparison centers} provides a summarized visualization of the probability of detection $P_D$ of significant differences observed in each center, represented by a color map. After eliminating the defective scans, certain centers show a lack of significant differences across all brain regions (falling into the first case described earlier). Conversely, centers exhibiting a constant $P_D$ across different regions belong to the second case discussed previously. Among the centers, the NYU center stands out as its results align more closely with the expected outcomes and are similar to those obtained when studying patients according to their condition. It is worth noting that some specific centers display high $PD$, i.e. $>$0.05 in almost all regions. The high frequency of significant differences observed corresponds to very limited sample sizes, with sizes smaller than 17 scans, in relation to the 'magical' number proposed in \cite{friston_ten_2012}.

\begin{figure}
    \centering
    \subfloat{
        \label{HC vs ASC with defec brain}
        \includegraphics[width=0.5\textwidth]{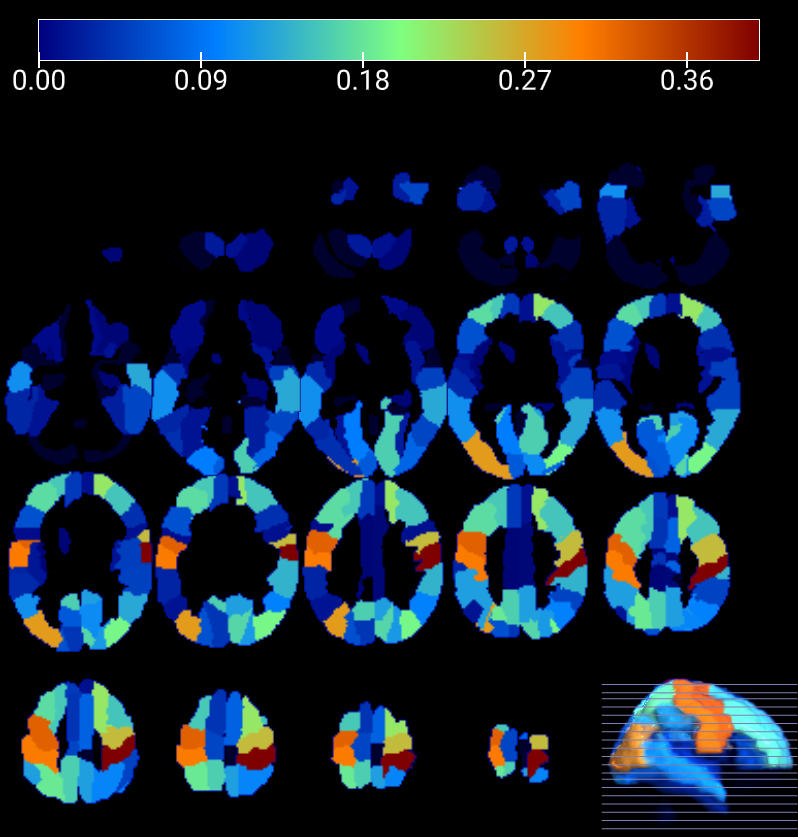}}    
    \subfloat{
        \label{HC vs HC with defec brain}
        \includegraphics[width=0.5\textwidth]{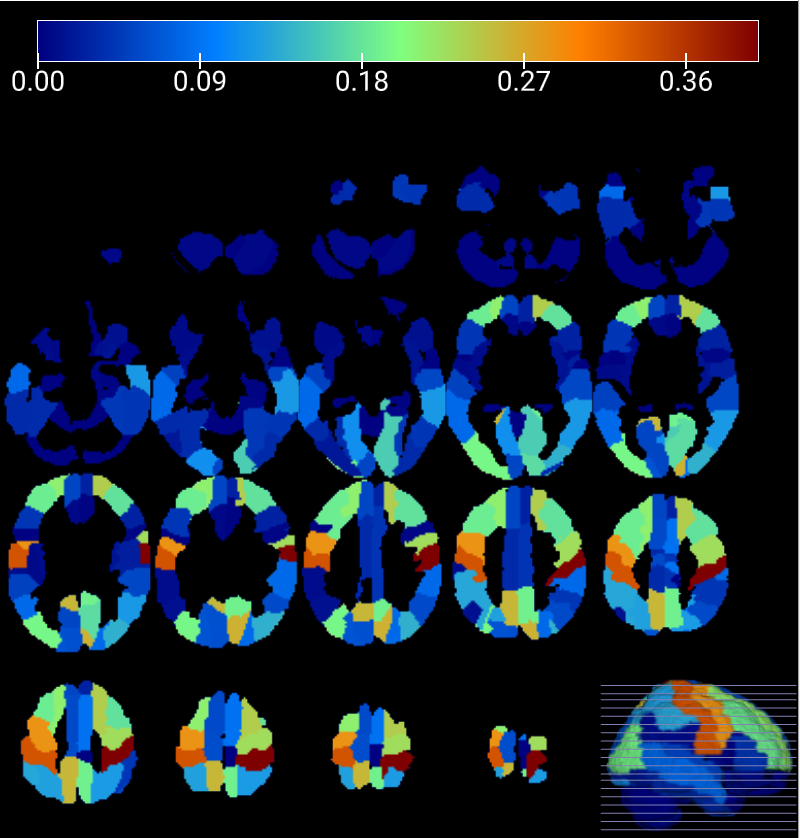}}

    \subfloat{
        \label{HC vs ASC without defec brain}
        \includegraphics[width=0.5\textwidth]{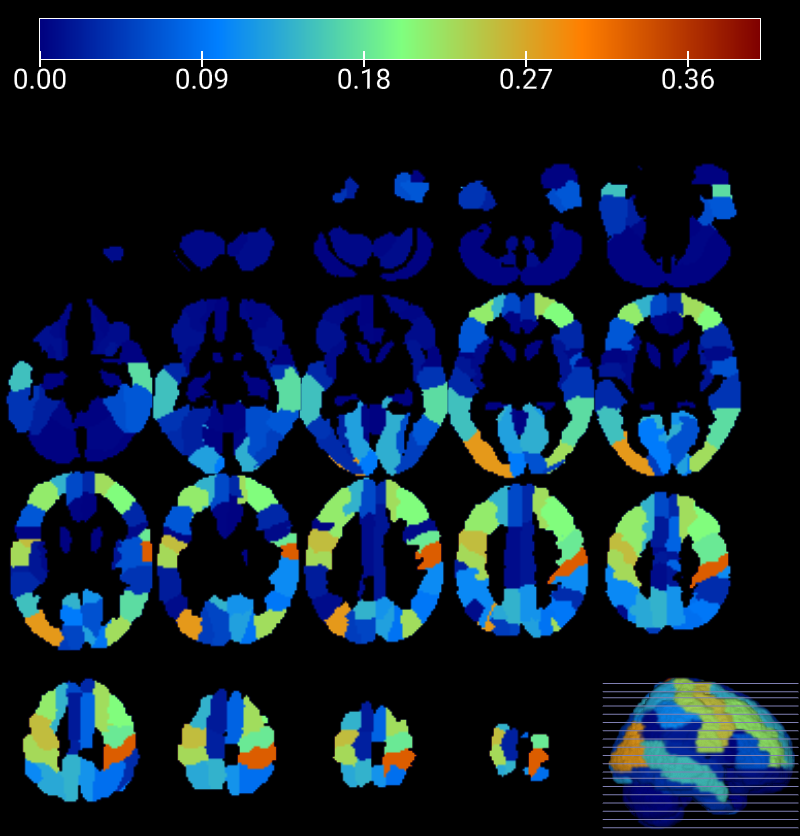}}    
    \subfloat{
        \label{HC vs HC without defec brain}
        \includegraphics[width=0.498\textwidth]{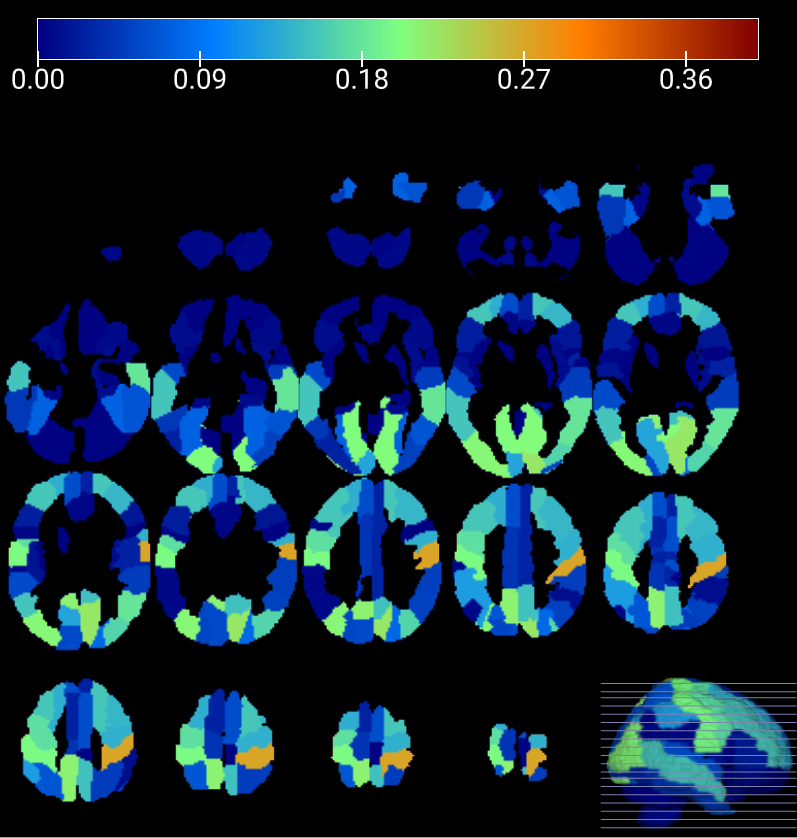}}

    \caption{All center study using the \ac{SAM} mapping method, highlighting the $P_D$ observed in each region during the comparisons \ac{HC} vs \ac{ASC} and \ac{HC} vs \ac{HC} from left to right, respectively. Above are the comparisons prior to the removal of flawed scans. Below, the same comparisons are presented after removing flawed scans.}
    \label{mosaico color reg}
\end{figure}

\begin{figure}
    \centering
    \subfloat{
        \label{HC vs ASC all scatter}
        \includegraphics[width=\textwidth]{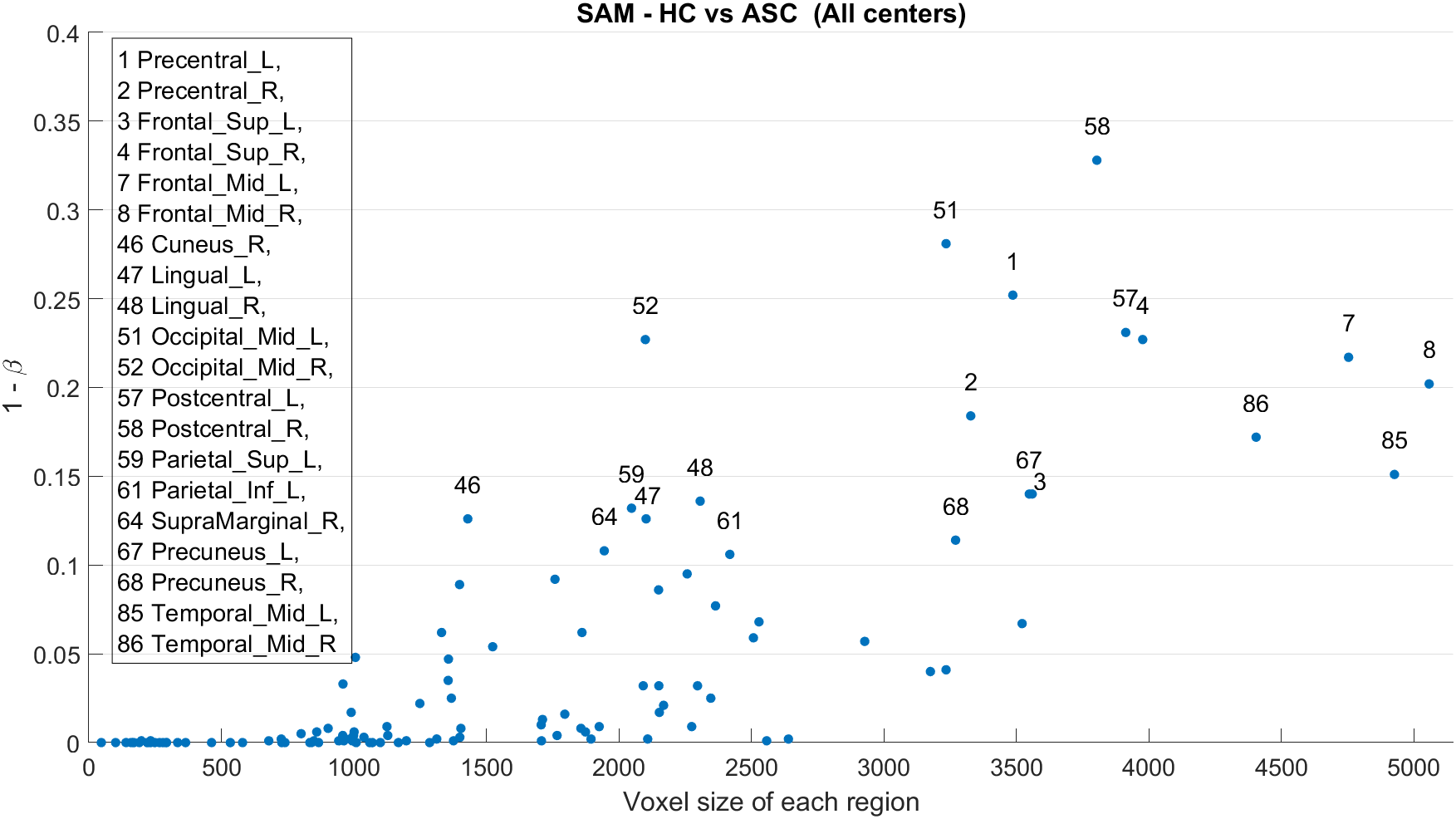}}
        
    \subfloat{
        \label{HC vs HC all scatter}
        \includegraphics[width=\textwidth]{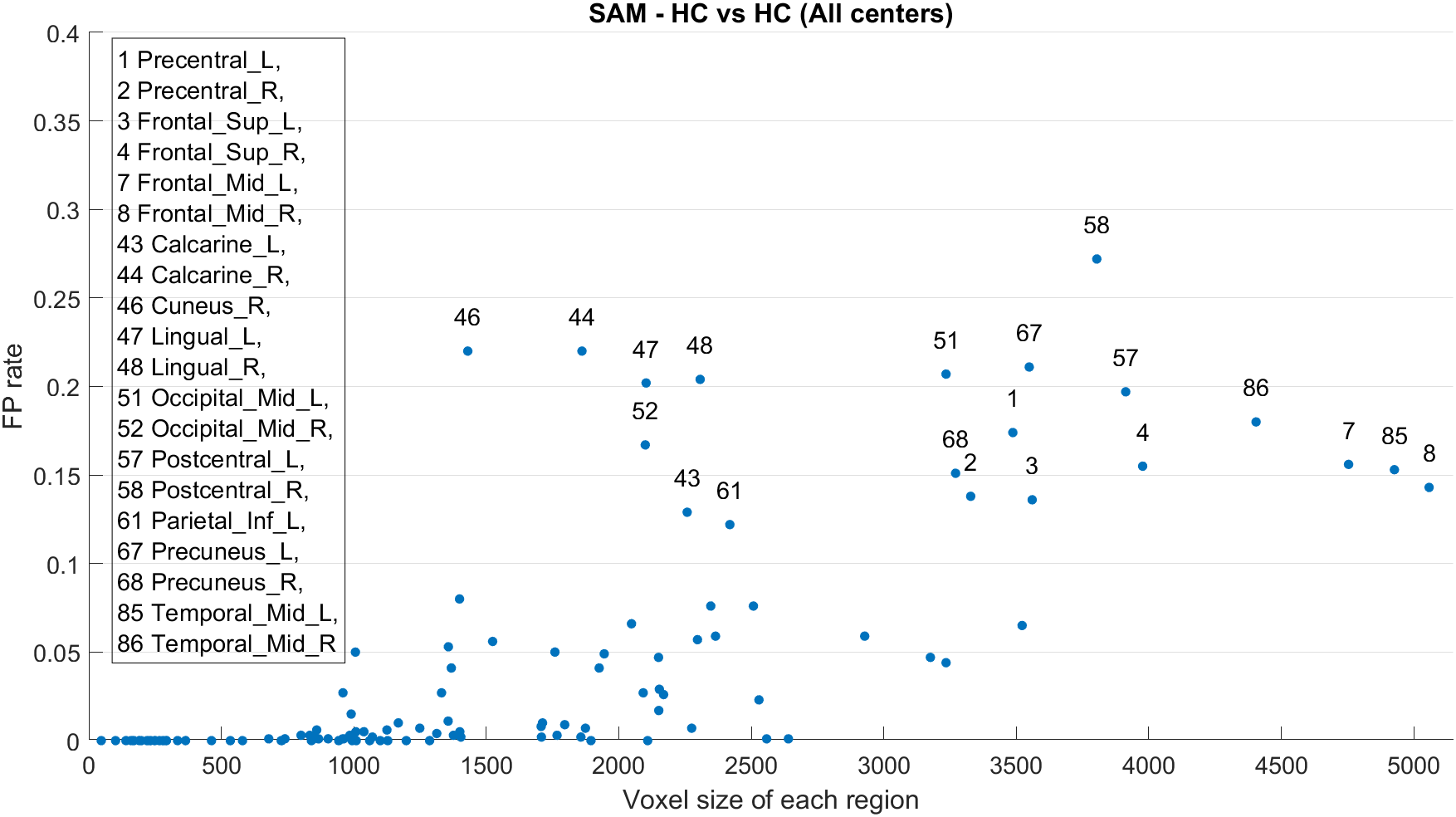}}
    \caption{Upper image shows estimated $P_D$ using SAM mapping method during the all centers comparison. Lower image shows FP rate in the same conditions. Each data point on the diagram represents a specific brain region, highlighting the most prominent ones.}
    \label{SAM_all_centers}
\end{figure}

Moreover, a new permutation test was conducted to study all the subjects according to their condition. The results, as shown in Figure \ref{mosaico color reg}, indicate that the regions exhibiting significant differences remain largely consistent in both comparisons. However, the frequency of these differences slightly changed after the elimination of the defective scans. Notably, regions such as Postcentral\_R and Occipital\_Mid\_L have emerged as significant in the updated results, as shown in Figure \ref{SAM_all_centers}. 

\begin{figure}
    \centering
    \subfloat[HC vs ASC mosaic]{
        \label{HC vs ASC mosaic diference}
        \includegraphics[width=0.5\textwidth]{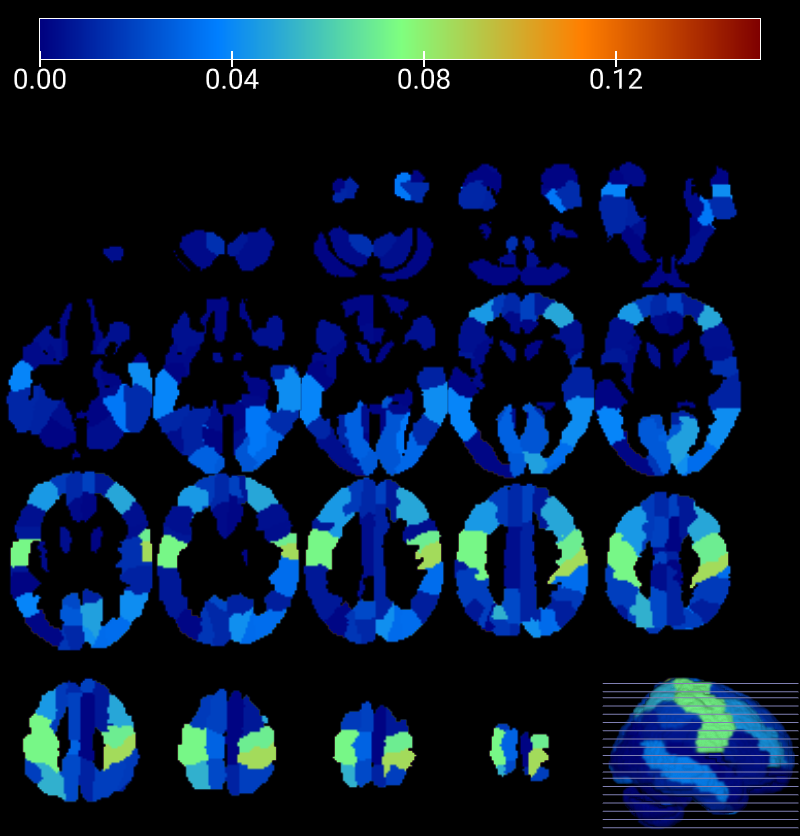}}    
    \subfloat[HC vs HC mosaic]{
        \label{HC vs HC mosaic diference}
        \includegraphics[width=0.5\textwidth]{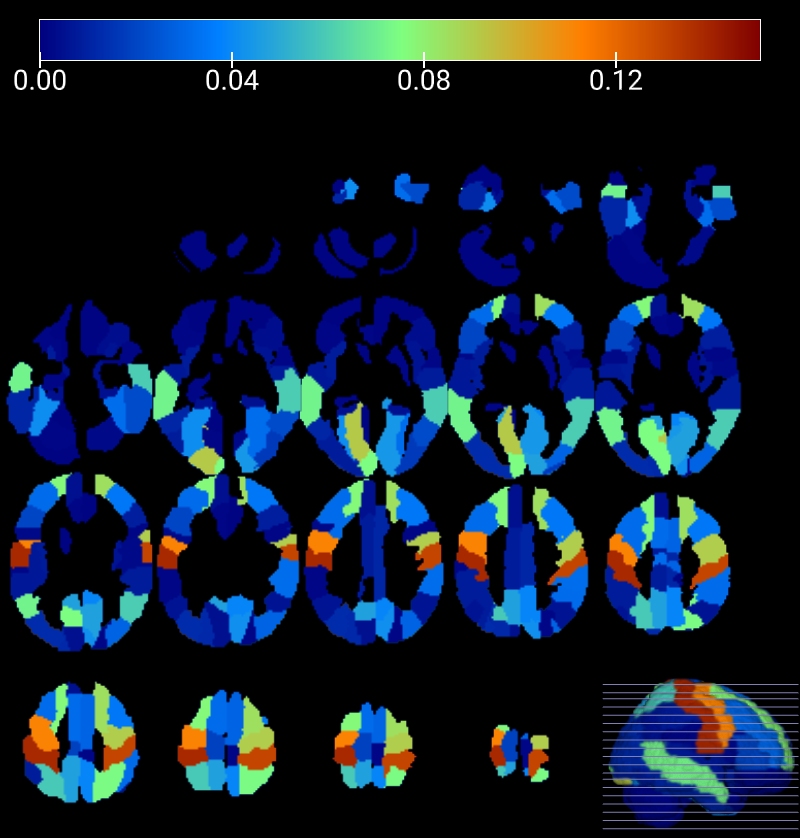}}
    \caption{Differences between the $P_D$ at each brain region including or excluding noisy data for \ac{HC} vs \ac{ASC} and \ac{HC} vs \ac{HC} comparisons. } 
    \label{mosaico diferences}
\end{figure}

Figure \ref{mosaico diferences} illustrates the observed difference in $P_D$ between the analysis conducted with the inclusion of defective scans and the analysis performed without them. This can be employed to assess the impact of noisy data on the detection of significant differences between study groups.

Based on the results obtained from the analysis by centers, we can conclude that centers characterized by a limited number of patients do not possess a sufficient sample size to yield meaningful and reliable insights. These centers provide overly optimistic or pessimistic maps of significance as shown in figure \ref{comparison centers}. In the case of the NYU center, where a substantial number of patients are available, it presents an opportunity for further in-depth study, offering the potential to generate more robust and informative outcomes.

\subsubsection{Study of NYU Center}

\begin{figure}
    \centering
    \subfloat[HC vs ASC NYU ]{
        \label{HC vs ASC NYU mosaico}
        \includegraphics[width=0.5\textwidth]{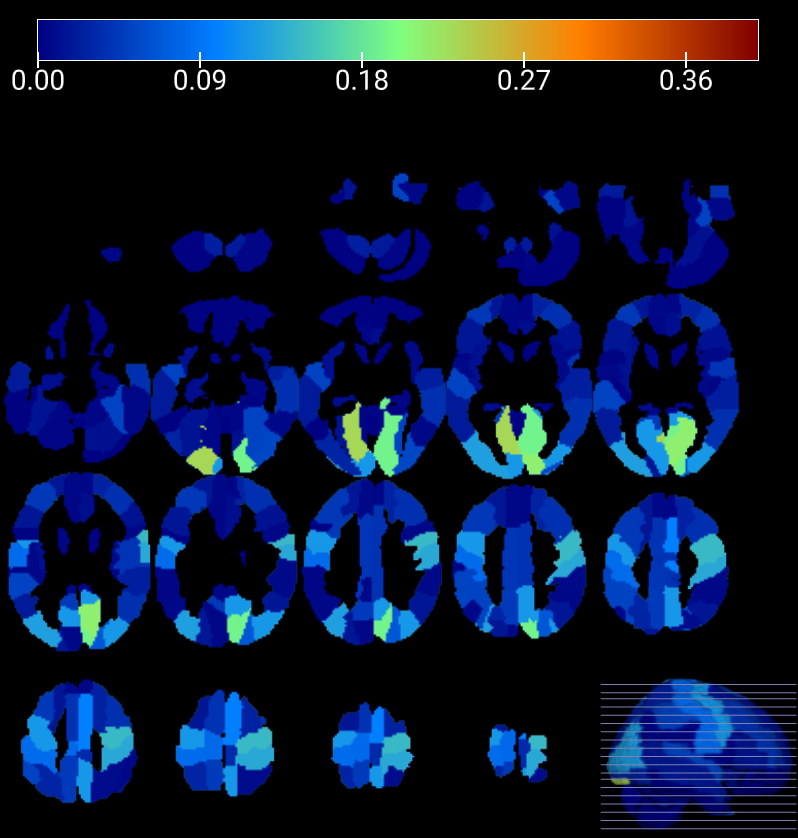}}    
    \subfloat[HC vs HC NYU ]{
        \label{HC vs HC NYU mosaico}
        \includegraphics[width=0.5\textwidth]{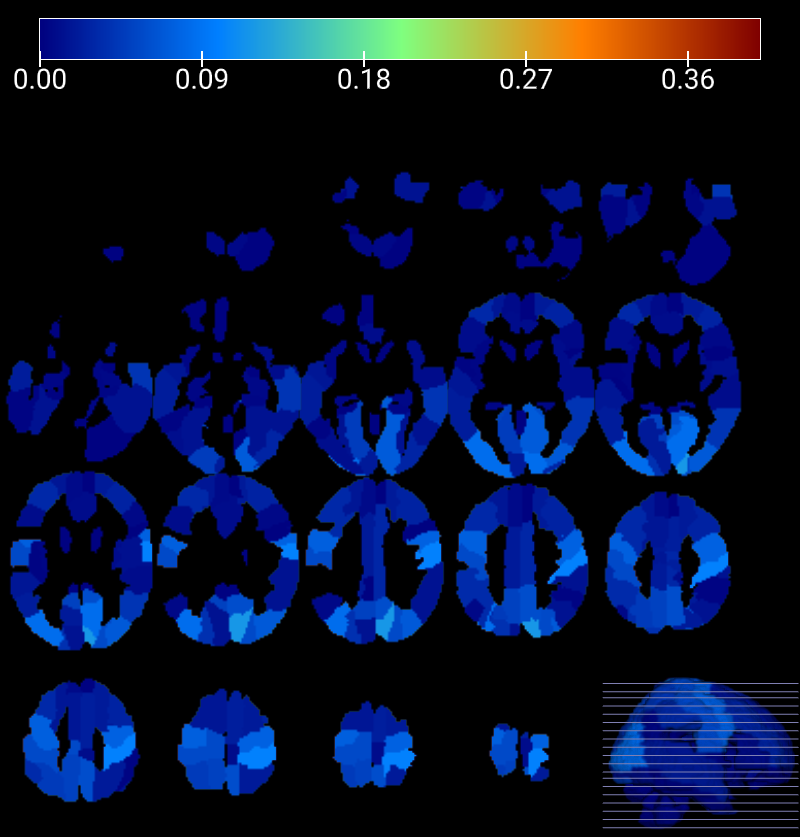}}

    \subfloat[]{
        \label{HC vs ASC NYU scatter}
        \includegraphics[width=0.5\textwidth]{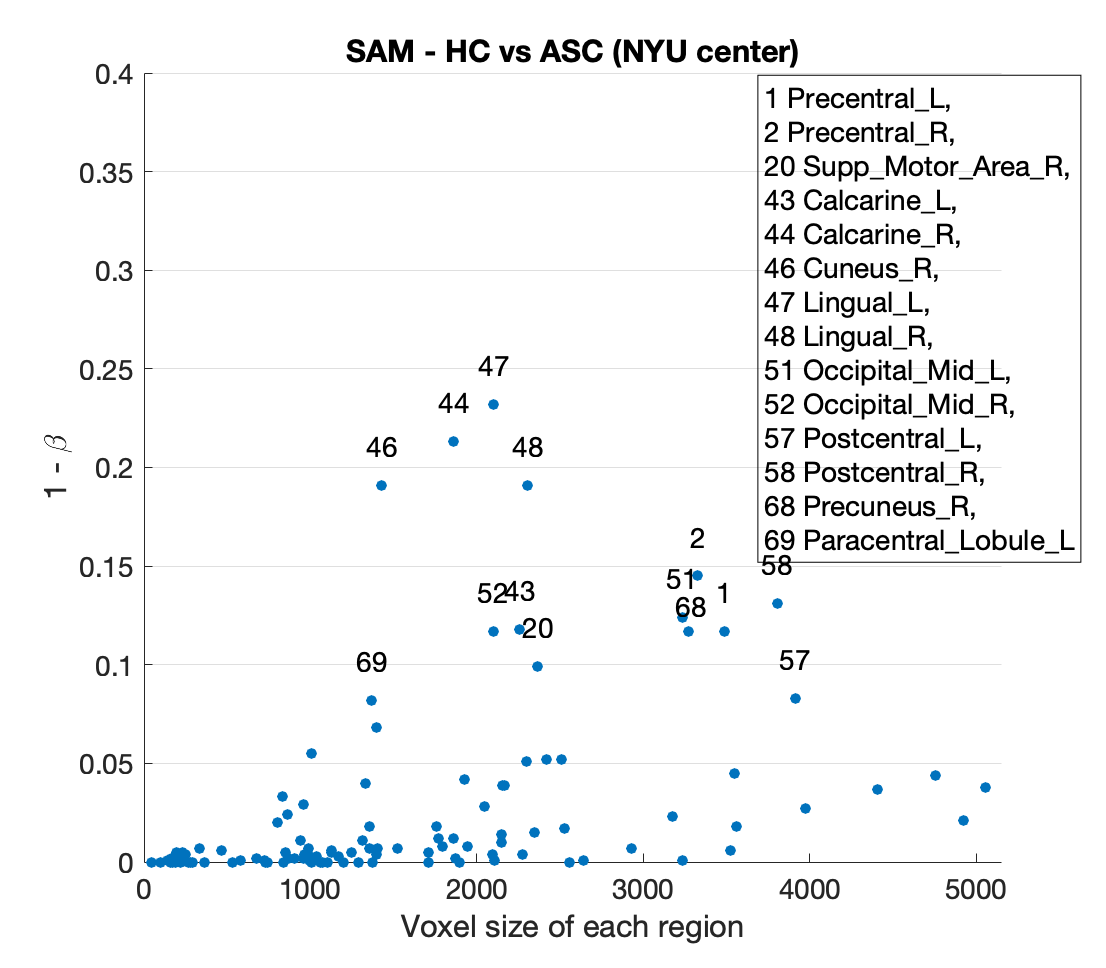}}    
    \subfloat[]{
        \label{HC vs HC NYU scatter}
        \includegraphics[width=0.5\textwidth]{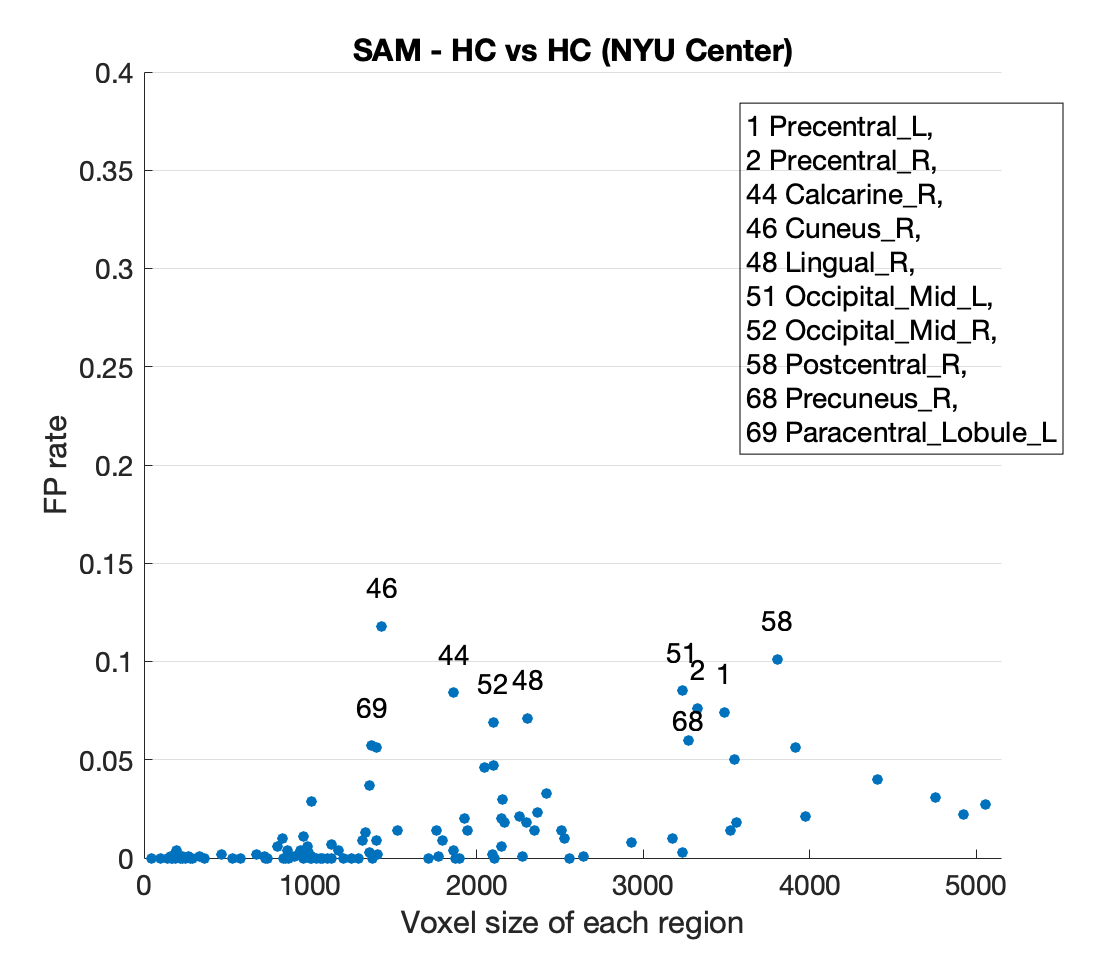}}
        
    \caption{Brain regions with significant differences in both comparisons (a) \ac{HC} vs \ac{ASC} and (b) \ac{HC} vs \ac{HC} using the \ac{SAM} permutation test after defective images removal in NYU. (c) Region voxel size vs. estimated $P_D$ in the \ac{HC} vs \ac{ASC} comparison using \ac{SAM}. (d) Region voxel size vs. FP rate in the \ac{HC} vs \ac{HC} comparison using \ac{SAM}. Each data point on the diagram represents a specific brain region, highlighting the most prominent ones.}
    \label{NYU center}
\end{figure}

The NYU center, with its large number of images, provides a robust dataset for focused studies. However, the results obtained from the permutation test in the NYU center reveal that the same brain regions consistently exhibit significant differences in both comparisons, while other regions consistently show no significant differences (Figure \ref{NYU center}). Notably, posterior brain regions such as Calcarine\_R, Cuneus\_R, Lingual\_L, and Lingual\_R stand out prominently. 

Considering the smooth changes in frequencies observed across different centers in figure \ref{comparison centers}, it is plausible that a relationship exists between the size or number of voxels comprising each region and its significance. On the left in Figure \ref{HC vs ASC NYU scatter}, we show a scatter plot comparing the voxel size of each region with the number of times it exhibits significant differences (statistical power if we assume an effect in the sample). A threshold of 0.05 has been set on the y-axis (significance level). The scatter plot suggests a relationship between these two variables, indicating that the power of the test increases as larger regions are considered. It can be inferred that regions below 1000 voxels are never significant. These findings could highlight the importance of considering multivariate approaches and the size of brain regions when analyzing significance, as larger regions provide more reliable and informative results in neuroimaging studies. 

Unfortunately, a similar effect on the rate of false positives (FPs) is observed in the right part of the same figure (Figure \ref{HC vs HC NYU scatter}). This observation indicates that there is indeed a trivial effect present in the entire dataset, with effects correlated in both groups. Upon closer examination of the results from the permutations test, it becomes apparent that the brain region with the most significant differences in both comparisons occurs approximately 240 times out of the 1000 permutations. The calculated $p-$values (average) for the brain regions in Table \ref{p_value_regions} were found to be higher than the predefined significance level of $\alpha = 0.05$ (see Table \ref{p_value_regions}). Therefore, besides the abnormal rate of FPs, it can be concluded that similar brain regions in the NYU center images exhibited non-significant differences in both comparisons. However, the surprisingly high rate of FPs in the HC vs. HC comparison affects the computation of p-values in the ASC vs. HC study, as shown at the bottom of the table. Assuming the $p-$values follow a normal distribution according to the empirical rule (68-95-99.7), the lack of evidence for rejecting the alternative hypothesis doesn't prove that the effect does not exist at the significance level (95\%).

\begin{table}
    \centering
    \caption{P-values calculated for the regions of the NYU center. Only the regions that exhibited a frequency of significant differences within the confidence interval are included.}
    \begin{tabular}{|c|c|c|c|}
    \hline
      Regions & $p-$value & Regions & $p-$value  \\
      \hline
      \hline
      Precentral\_L & 0.3337 & Occipital\_Mid\_R & 0.3566 \\
      \hline
      Precentral\_R & 0.4096 & Occipital\_Inf\_R & 0.4216 \\
      \hline
      Supp\_Motor\_R & 0.4296 & Fusiform\_R & 0.4316 \\
      \hline
      Calcarine\_L & 0.4945 & Postcentral\_L & 0.3177 \\
      \hline
      Calcarine\_R & 0.4705 & Postcentral\_R & 0.3037 \\
      \hline
      Cuneus\_R & 0.3317 & Parietal\_Inf\_L & 0.3656 \\
      \hline
      Lingual\_L & 0.4575 & Precuneus\_L & 0.2757 \\
      \hline
      Lingual\_R & 0.4146 & Precuneus\_R & 0.3776 \\
      \hline
      Occipital\_Sup\_R & 0.3347 & Paracentral\_Lob\_L & 0.3576 \\
      \hline
      Occipital\_Mid\_L & 0.3487 & Temporal\_Sup\_L & 0.4046 \\
      \hline
    \end{tabular}
    \includegraphics[width=\textwidth]{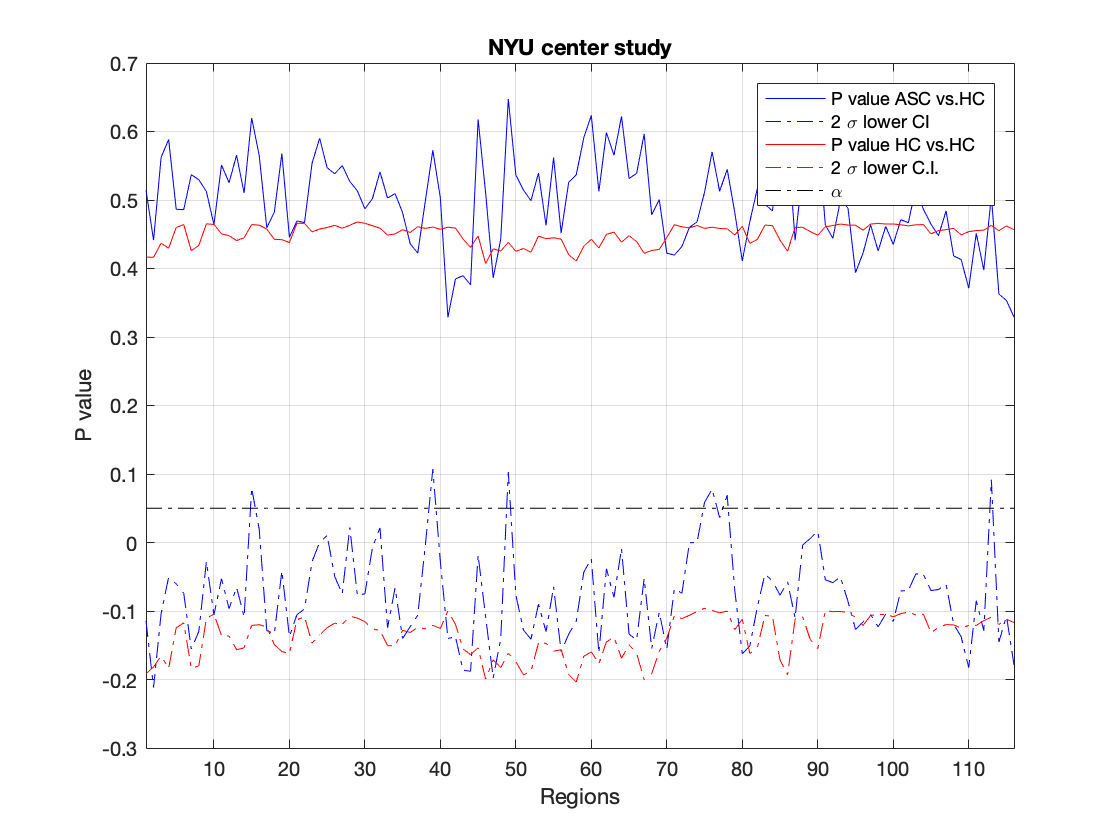}  
    \label{p_value_regions}  
\end{table}

\subsection{Analysis with SPM}

The initial study was conducted by grouping all the images based on patient condition, disregarding the centers to which they belong. The images were divided into two groups: \ac{ASC} and \ac{HC}. Two variables were studied: the frequency at which different voxels were present in each region during the 1000 permutations, and the number of significant voxels in each region. These two measures account for the significance of a region. 

\begin{figure}
    \centering
    \subfloat{
    \label{SAM_vs_SPM_color}
    \includegraphics[width=0.6\textwidth]{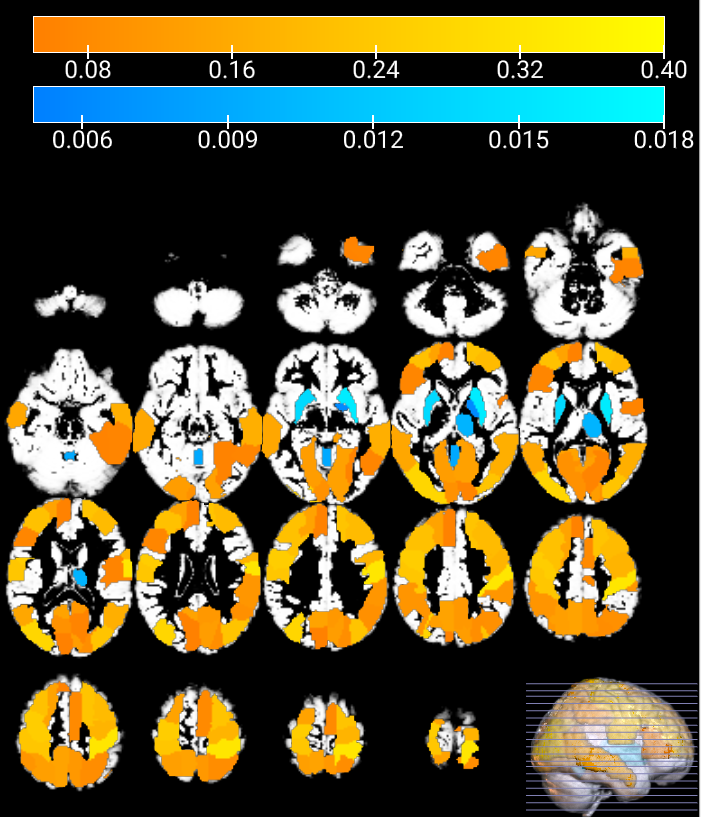}}

    \subfloat{
    \label{SAM_vs_SPM_plot}
    \includegraphics[width=\textwidth]{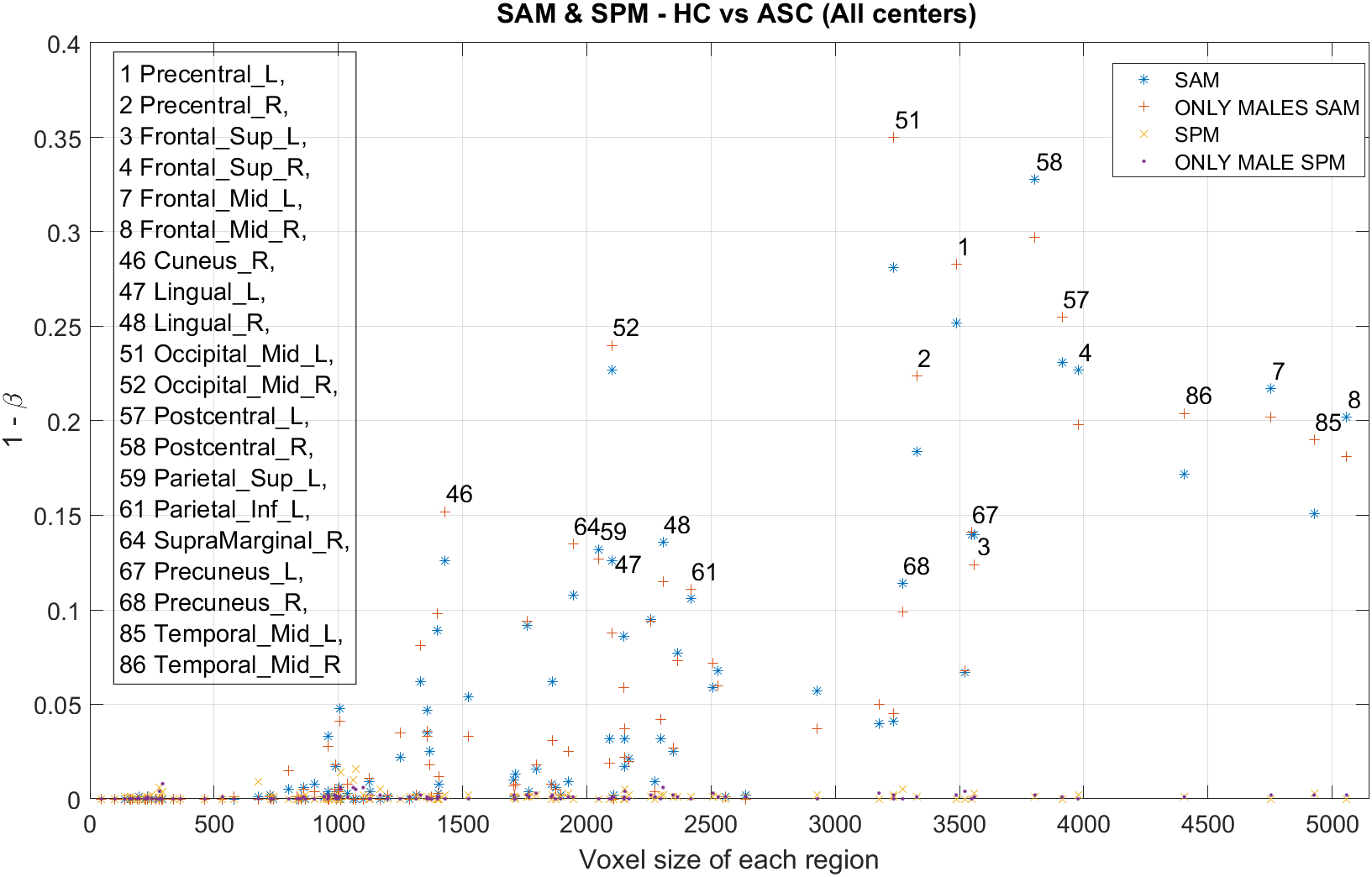}}
    \caption{\ac{SAM} and \ac{SPM} results for \ac{HC} vs \ac{ASC} all centers comparison. In the upper image, it is represented in orange the $P_D$ at each brain region detected by \ac{SAM} statistical method and in blue the $P_D$ detected by \ac{SPM}. In the lower image, the SAM and SPM comparison is presented, including their respective analyses with male subjects only.}
    \label{SAM_SPM_all}
\end{figure}

At the bottom of Figure \ref{SAM_SPM_all}, we computed the $P_D$ of the SPM voxelwise statistical framework and compared it to the one obtained with the SAM method. As clearly shown in the figure, SPM did not provide the nominal value of 0.05, indicating an overly super-conservative method. This behaviour is indeed persistent in the studies with larger samples sizes as shown in the next section. Following these results, the $p-$value analyses did not reveal any evidence to reject the null hypothesis at the significance level.

\subsubsection{Study by center}

The \ac{SPM} analysis failed in Group I (CALTECH, KKI, LEUVEN 1, LEUVEN 2, MAX MUN, OLIN, SBL, SDSU, STADFORD, TRINITY, USM)  centers due to a limited number of samples available for analysis. On the other hand, the NYU center, which has the largest number of images, demonstrates minimal voxels with differences and a significantly lower frequency of regions showing significant differences. In fact, the regions exhibiting differences in the NYU center fall well below the threshold of 50, which was previously established as the nominal value or the expected value of FPs under the null hypothesis. This observation holds true for both the study by condition and the study by centers using \ac{SPM}.

\begin{figure}[t]
    \centering
    \subfloat[]{
        \label{Voxels with atlas}
        \includegraphics[scale=0.32]{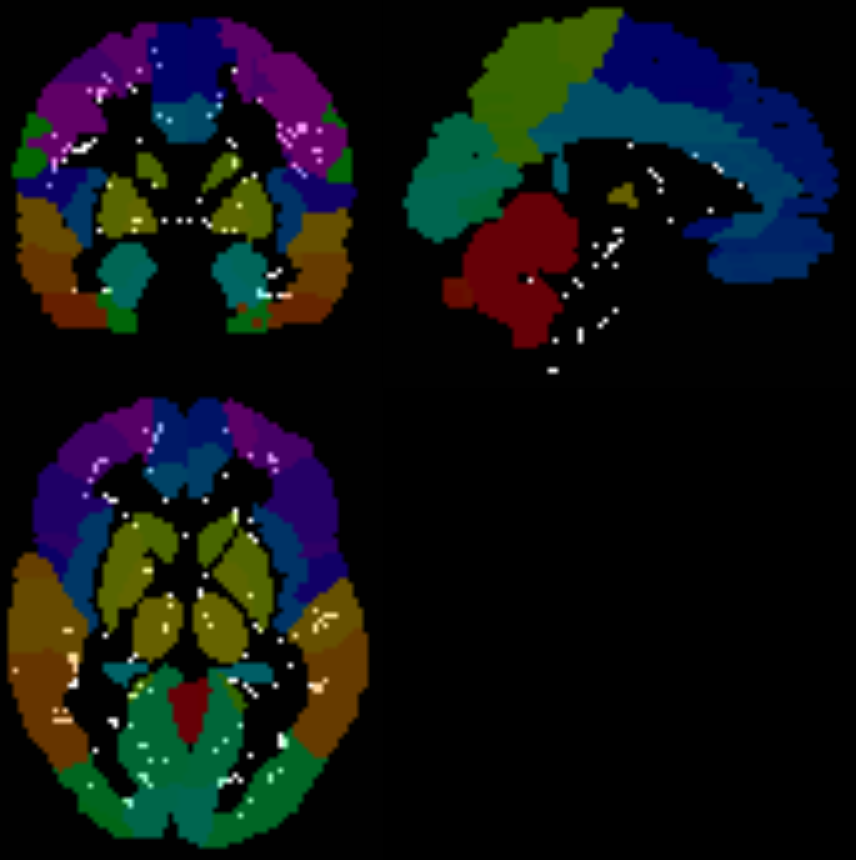}}    
    \subfloat[]{
        \label{voxels with MRI}
        \includegraphics[scale=0.32]{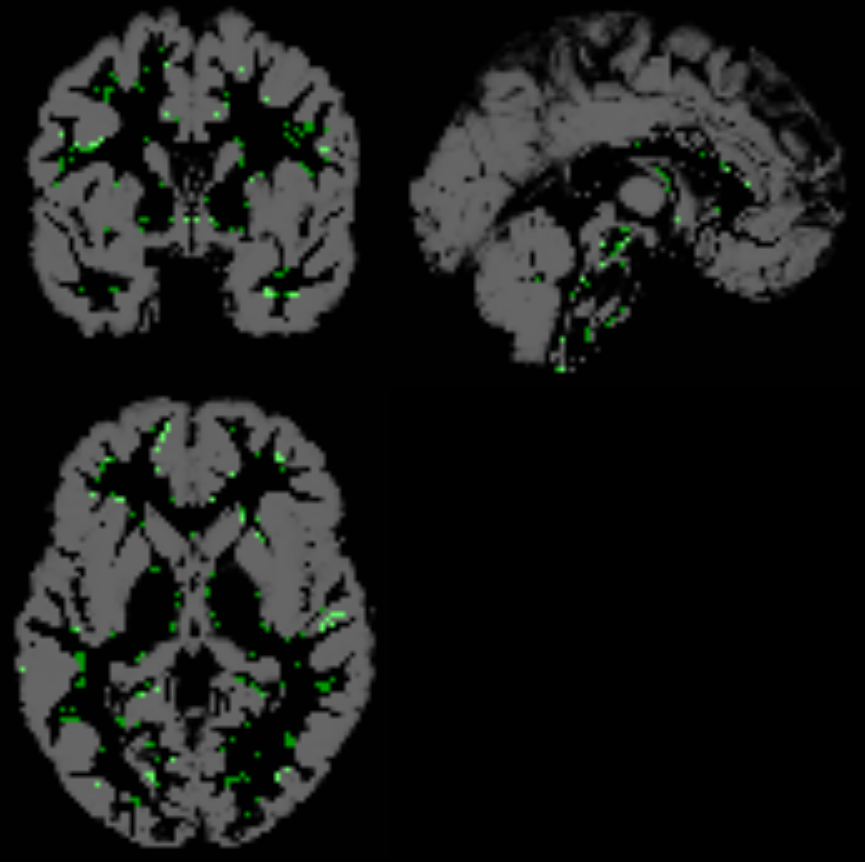}}
    \caption{\ac{SPM} t-map in \ac{ASC} vs \ac{HC} comparison of OLIN center, compared to the AAL atlas (a), and compared to the \ac{GM} map (b).}
    \label{voxel + atlas y mri}
\end{figure}

On the contrary, some limited samples exhibit a higher number of voxels showing differences after the contrast analysis. This is clearly an undesirable behaviour taking into account the outcome with larger sample sizes. Taking the OLIN center as an example, we can observe the voxels that have been identified as different after the permutation test using \ac{SPM} in Figure \ref{voxel + atlas y mri}, superimposed on the image provided by the \ac{AAL} atlas. These voxels, located near the edges of different regions, can be attributed to registration artifacts, which can arise from anatomical variations among subjects or discrepancies between the atlas-defined regions and the detected regions during image registration. Also, some differences are concentrated in the brain stem, as it is shown in the same figure.

\begin{figure} 
    \centering
    \subfloat[ASC vs HC voxels comparison of Group I centers with SPM]{
        \label{voxel ASC vs HC group I centers}
        \includegraphics[width=\textwidth]{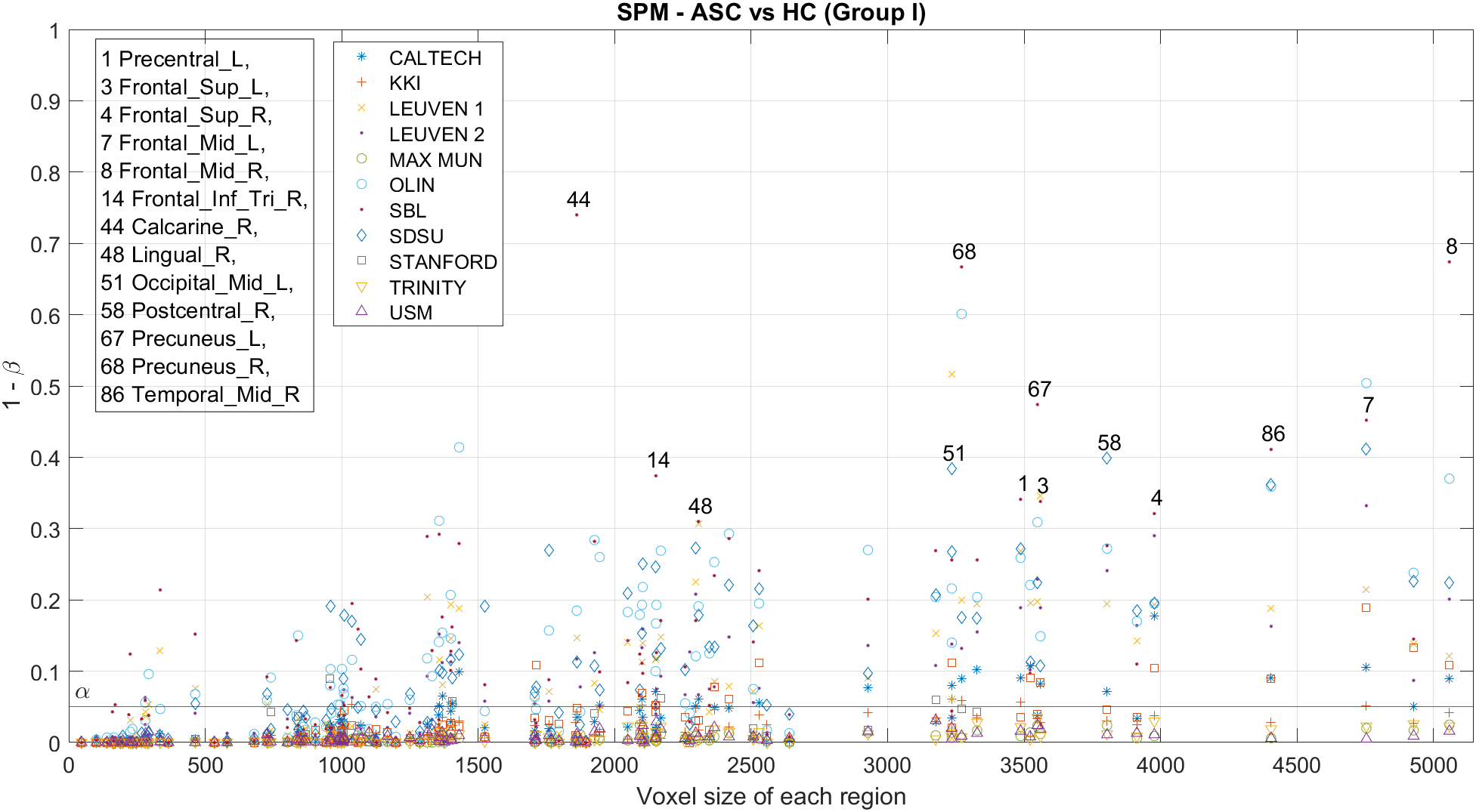}}   
        
    \subfloat[HC vs HC voxels comparison of Group I centers with SPM]{
        \label{voxel HC vs HC group I centers}
        \includegraphics[width=\textwidth]{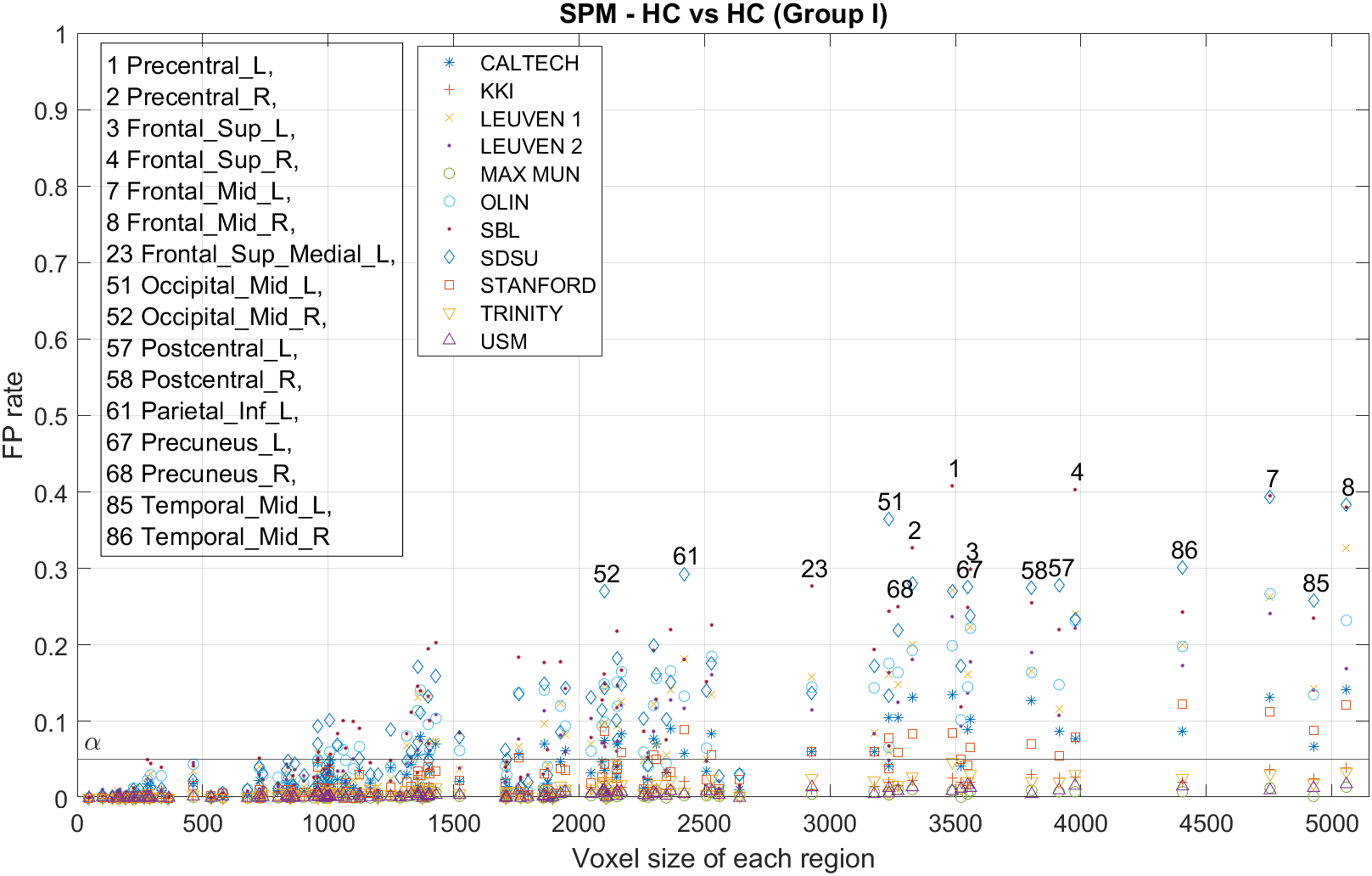}}
    \caption{Estimated $P_D$ (upper) and FP rate (lower) with SPM during the Group I centers analysis. Each point on the diagram represents a specific brain region, highlighting the most prominent ones.}
    \label{voxel_groupI_centers}
\end{figure}

Figure \ref{voxel_groupI_centers} reveals the frequency of significant voxels across various brain regions following the SPM contrast for centers with limited sample sizes (Group I). Notably, these centers exhibit relatively high voxel frequencies and disparities between regions, along with numerous algorithmic issues during the analysis. Consequently, it can be concluded that the large number of voxels detected in these centers is not attributable to the identification of a specific region marking differences between individuals with autism and healthy controls. Instead, it likely stems from center-specific characteristics, such as sample realization, confounding effects, etc.

\begin{figure}
    \centering
    \subfloat[ASC vs HC voxels comparison of Group II centers with SPM]{
        \label{voxel ASC vs HC group II centers}
        \includegraphics[width=\textwidth]{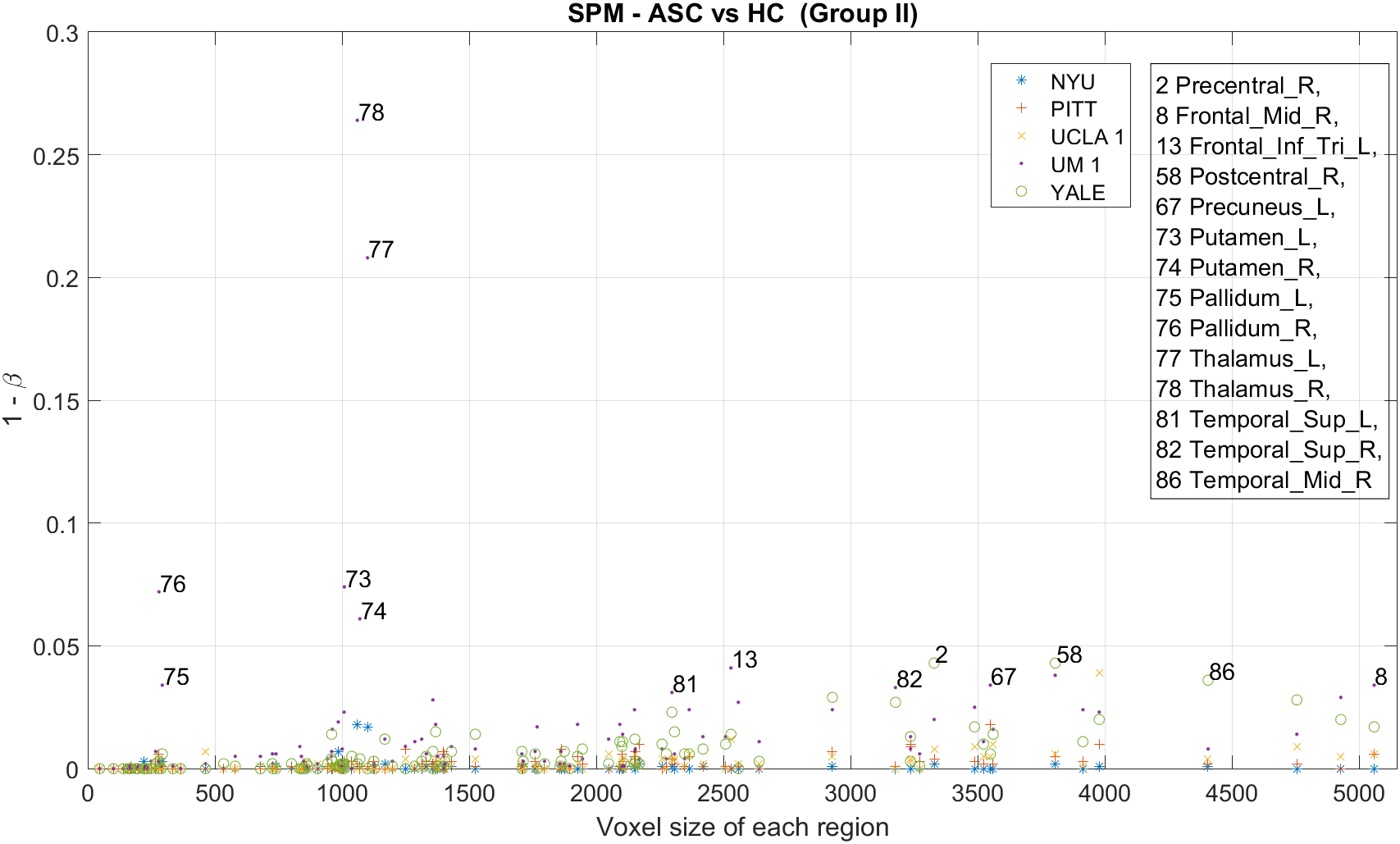}}   
        
    \subfloat[HC vs HC voxels comparison of Group II centers with SPM]{
        \label{voxel HC vs HC group II centers}
        \includegraphics[width=\textwidth]{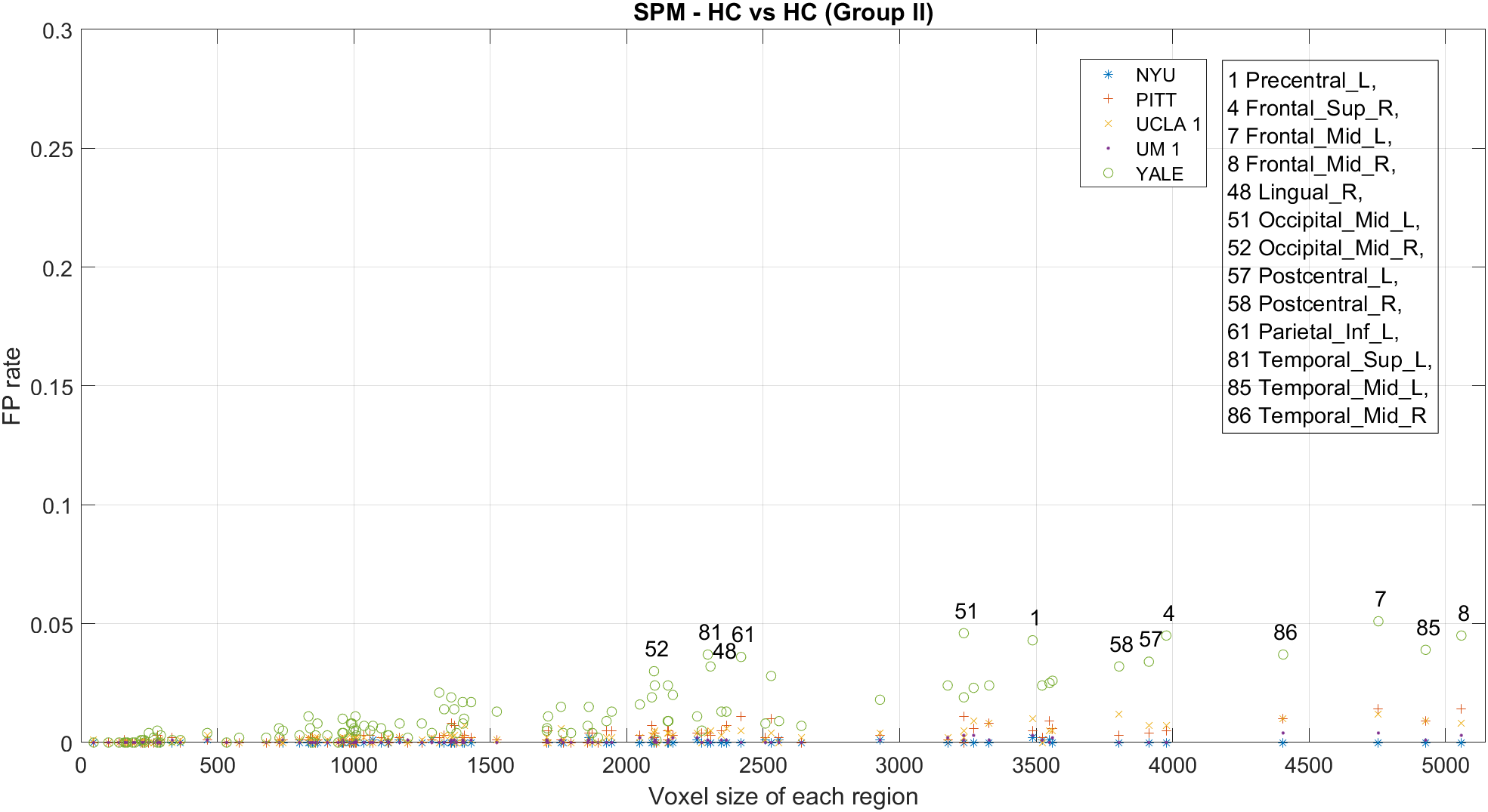}}
    \caption{Estimated $P_D$ (upper) and FP rate (lower) with SPM during the Group II centers analysis. Each point on the diagram represents a specific brain region, highlighting the most prominent ones.}
    \label{voxel_groupII_centers}
\end{figure}

Figure \ref{voxel_groupII_centers} show the results on centers larger sample sizes (Group II: NYU, PITT, UCLA\_1, UM\_1, and YALE). Although there are regions that exhibit a substantial number of voxels, these completely vanish in the analysis across all centers.

After analyzing the images with \ac{SPM}, both by condition and by center, it can be concluded that \ac{SPM} does not identify any region or voxel that exhibits statistically significant differences at $\alpha=0.05$. Therefore, based on the results obtained, it could be inferred that there are no specific brain regions or voxels that can be considered reliable indicators for distinguishing between individuals with autism and neurotypical individuals. Taking into account all the issues and artifacts found in the analyzed datasets, this is rather a consequence of the super conservative behavior of voxelwise statistical inferences in the SPM method.

\section{Discussion}

Throughout this discussion, the results ascertained from performing image analysis on the \ac{ABIDE I} database using each of the selected mapping methods will be scrutinized to determine the extent of their resemblance or divergence. Emphasis will be placed on two key aspects: (1) the study encompassing patient conditions, and (2) the examination of individual centers within the \ac{ABIDE I} database, accounting for their respective images. Subsequently, a comparative analysis will be conducted to juxtapose our findings with the existing body of scientific literature. This examination aims to assess the extent of concurrence between our results and the previously documented research in the field. 

\subsubsection{SAM and small sample sizes}

After analyzing each center using \ac{SAM}, it was observed that some centers had completely null results, meaning that their regions did not exhibit any significant differences during the 1000 permutations, not even due to randomness. Conversely, there were centers that showed regions with consistently similar levels of significance across the 1000 permutations, illustrated in Figure \ref{comparison centers}. 

The primary reason for these results can be attributed to the insufficient number of samples available at each center, which provides us with a super-conservative upper bound. Most centers had relatively small databases, with fewer than 20 samples for each group, except for a few centers. Additionally, the removal of defective brain images further decreased the sample size. Consequently, during the permutation test, the total number of samples ranged from 8 to 20, evenly distributed between the two groups. Such a limited sample size could strongly influence the atypical results observed in these cases, as \ac{SAM} assumes some degree of uncertainty in the data and applies statistical corrections based on the assumption of randomness.

Moreover, the presence of artifacts and confounding effects in the sample realization provides ``actual'' classes and subclusters in the population, resulting in heterogeneous samples. Consequently, with a small number of samples, null empirical errors can be reached, compensating for the ultra-conservative upper bounds, even in low-dimensional scenarios. Therefore, SAM's methodology only reflects the actual classes that conform to the groups.

Considering previous studies indicating brain differences related to sex within the autism spectrum\cite{gorriz_machine_2019}, the  differentiation of male subjects from female subjects was studied. Focusing solely on male patients due to their larger representation in the dataset, the results obtained did not deviate from previous findings, suggesting that the sex of the patient did not significantly influence the outcomes, as shown in Figure \ref{SAM_vs_SPM_plot}. This lack of impact could be attributed to the small proportion of female patients in the dataset, limiting their statistical relevance.

\subsection{Study by condition}

\ac{SAM} analysis revealed the presence of brain regions with a notable number of significant differences for both comparisons (\ac{HC} vs \ac{ASC} and \ac{HC} vs \ac{HC}), even after excluding defective brains. However, the observed differences did not reach a sufficiently high level compared to the number of permutations conducted, thereby precluding the identification of any decisive regions as shown in the $p$-value analysis. Indeed, upon fixing the significance level and calculating the $p$-value for each region, it was determined that none of them exhibited statistically significant differences, although in this case we prefer to claim that there is no sufficient evidence in support of the alternative hypothesis. Similarly, the analysis performed with \ac{SPM} demonstrated a scarcity of significant voxels in the T-maps. The frequency of regions exhibiting differences at voxel level during the 1000 permutations was even below the nominal level (50), affirming the absence of statistically significant differences in any particular region using this over-conservative method. 

Figure \ref{SAM_SPM_all} provides a comparison of \ac{SAM} and \ac{SPM} results, visualizing regions identified as different by \ac{SAM} along with their corresponding frequency values using an orange color map, and highlighting the number of voxels that manifested differences after the \ac{SPM} contrast using a blue color map. \ac{SAM} hightlights regions as Postcentral\_R, Postcentral\_L, Occipital\_Mid\_L, Occipital\_Mid\_R, Precentral\_L, among others. Meanwhile, \ac{SPM} highlights regions as Putamen\_L, Putamen\_R, Thalamus\_R, Vermis\_4\_5, among others. Notably, the highest concentration of voxels detected by \ac{SPM} was observed in regions where \ac{SAM} indicated no or minimal differences, while regions exhibiting more significant differences according to \ac{SAM} showed scarce significant voxels. Consequently, both methods did not individually establish statistically significant differences but contradicted each other in identifying regions with great disparities.

\subsection{Study by acquisition sites}

Both \ac{SAM} and \ac{SPM} methods revealed the need to differentiate the NYU center from the remaining centers in the study. In both methods, some centers have regions with significantly higher frequency or voxel count compared to NYU or the study conducted based on the subjects' condition. However, it is highly likely that these results, particularly in \ac{SAM} where the values among centers were similar, stem from the limited number of images available for each center. The limited number of images, regardless of the mapping method employed, contributes to errors that compromise the reliability of the obtained results. Conversely, when the images from all centers are aggregated and analyzed as a whole (study by condition), the differences found between groups dilutes. 

Regarding the NYU center, both \ac{SAM} and \ac{SPM} mapping methods exhibit remarkably similar results compared to the results obtained from the study conducted based on the subjects' condition. This can be attributed to the fact that NYU contributes the largest number of samples, making it the most suitable center for in-depth examination among the 20 centers comprising the \ac{ABIDE I} database.

In \ac{SAM}, the NYU center demonstrates numerous brain regions that exhibit differences exceeding 100 occurrences during the 1000 permutations. Notably, during the \ac{HC} vs \ac{ASC} comparison, which is of particular interest, the regions Calcarine\_R, Cuneus\_R, Lingual\_L, and Lingual\_R emerge as the regions with the most significant differences, with two of them exceeding 200 out of 1000 occurrences. However, \ac{SPM} analysis reveals a minimal number of voxels with differences after the contrast in the NYU center. The regions Hippocampus\_R, Thalamus\_L, and Thalamus\_R exhibit the highest concentration of these differing voxels, while \ac{SAM} either indicates no differences or minimal distinctions in these areas. Consequently, it can be concluded that both mapping methods contradict each other by identifying entirely distinct regions as positive. Nevertheless, both approaches reach the same statistical conclusion from the analysis of $p$-values.

\begin{figure}
    \centering
    \includegraphics[width=0.6\textwidth]{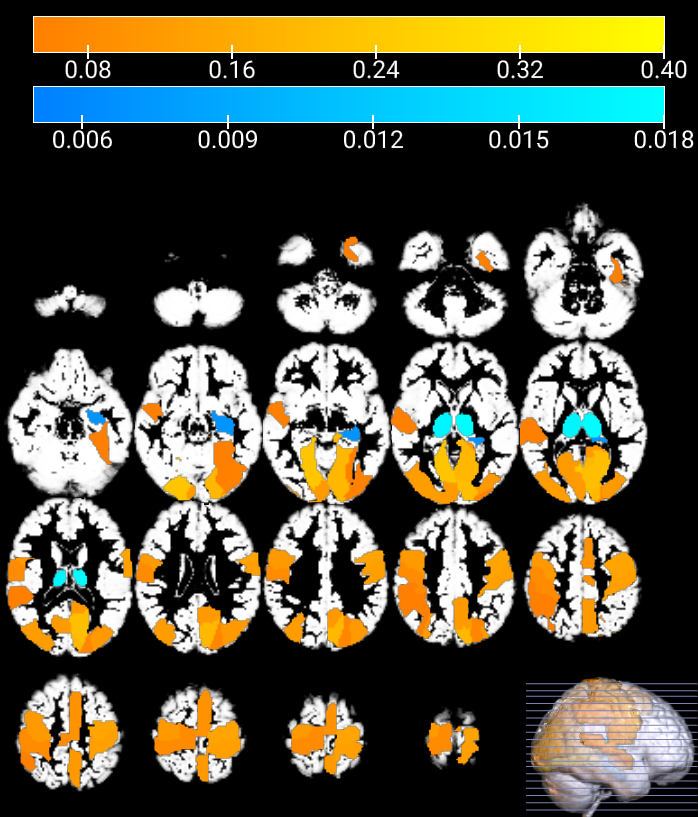}
    \caption{Comparison of \ac{SAM} and \ac{SPM} results for \ac{HC} vs \ac{ASC} comparison at NYU center. In orange, the differences in brain regions detected by \ac{SAM} statistical method. In blue, the differences detected by \ac{SPM}.}
    \label{SAM_SPM_NYU}
\end{figure}

Figure \ref{SAM_SPM_NYU} visually presents the comparison between \ac{SAM} and \ac{SPM} results, illustrating the regions identified by \ac{SAM} as having significant differences along with their corresponding frequency denoted by an orange color map, as well as the most significant voxels detected by \ac{SPM}, denoted by a blue color map.

This finding provides evidence that the examination of centers does not reveal any regions or voxels that exhibit statistically significant differences between individuals with \ac{ASC} and healthy controls. The absence of significant findings can be attributed to two factors: firstly, the limited availability of samples in the majority of centers, which compromises the statistical power of the analysis; and secondly, the regions under investigation do not demonstrate a substantial number of differences relative to the permutations conducted. Additionally, it is noteworthy that each statistical method identifies disparate regions as significant, despite employing the same set of images.

\subsection{Final remarks on the ABIDE dataset}
The findings from the analysis of \ac{ABIDE I} database images based on the subjects' condition indicate the absence of any specific brain region that can be deemed crucial in distinguishing individuals with \ac{ASC} from healthy controls. The lack of statistically significant differences suggests that a comprehensive differentiation between the two groups cannot be achieved solely through this approach. 

One possible explanation for these outcomes is the amalgamation of images from various centers within the \ac{ABIDE I} database, resulting in a heterogeneous multicenter database. Autism is inherently characterized by its heterogeneity, encompassing diverse symptoms and characteristics, hence its classification as a spectrum disorder \cite{lombardo_big_2019, masi_overview_2017, szatmari_heterogeneity_1999}. Consequently, attempting to study such a complex disorder using a database that comprises multiple centers with distinct imaging protocols, equipment, and laboratory settings, among other variables, may yield inconsistent results\cite{katuwal16}. To address this, the utilization of larger databases characterized by uniformity in scanner type and imaging procedures would be advantageous. By ensuring uniformity and control over extraneous factors unrelated to the disorder, the potential for false results arising from these confounding influences can be minimized. Therefore, it is advisable to avoid pursuing a multicenter approach in neuroimaging research when studying disorders as heterogeneous as autism without carefully controlling confounding variables, sample size, imaging protocols and other sources of variability not related to the disease.

\subsection{Comparison with the extant literature}

Currently, numerous studies have investigated the application of deep learning techniques using the \ac{ABIDE I} database to classify autism, wherein they have identified specific brain regions that exhibit notable correlations. These regions often manifest in the posterior section of the brain, including the Occipital Pole, Precuneus Cortex, Intracalcarine Cortex, Left Lingual Gyrus, and others \cite{bi_classification_2018, heinsfeld_identification_2017, yang_functional_2019, nielsen_multisite_2013, just_cortical_2004, epalle_multi-atlas_2021}. These areas hold potential significance in our image comparison study. Focusing solely on the \ac{HC} vs \ac{ASC} comparison, the \ac{SAM} analysis has revealed that the regions displaying a greater frequency of significant differences, both in the study based on the patients' condition and in the study conducted at the NYU center (yielding the most favorable outcomes), are predominantly located in the posterior regions of the brain. Noteworthy areas include Occipital\_Mid\_L, Calcarine\_R, Cuneus\_R, Lingual\_L, Lingual\_R, among others. Additionally, the Postcentral region stands out in the SAM analysis and in certain publications. Thus, these areas may align with those exhibiting robust correlations in previous studies. However, the \ac{SPM} analysis reveals that the majority of differences, or areas with the highest voxel concentration after contrast, are situated in the central and inner regions of the brain. This pattern is observed both in the analysis based on the patients' condition and in the study encompassing centers that encountered no issues during the analysis. Noteworthy regions encompass Putamen\_L, Putamen\_R, Thalamus\_L, Thalamus\_R, Frontal\_Sup\_R, Temporal\_Mid\_R, among others. Figures \ref{SAM_SPM_all} and \ref{SAM_SPM_NYU} prominently depict the regions of interest identified by both mapping methods, providing a clear visual representation.

\ac{SAM}, in contrast to \ac{SPM}, produces results that exhibit greater resemblance to the findings reported in existing literature, indicating overlapping regions of significance within the brain. Although these results may not meet the threshold for statistical significance, they contribute to narrowing down the search for key brain regions that can reliably differentiate between individuals with autism and neurotypical individuals.

It is noteworthy that certain studies, such as \cite{rakic_improving_2020, kong_classification_2019, ms_darkasdnet_2021}, have reported classification accuracies exceeding 80\% when utilizing images from the \ac{ABIDE I} database, with some works estimating the classification performance of the proposed methods to 100\%, as demonstrated in \cite{selcuk_nogay_diagnostic_2023}. However, our present research indicates that achieving such a high level of accuracy in autism classification with the \ac{ABIDE I} database is highly improbable, primarily because there is a lack of distinct structural brain patterns that can serve as clear differentiator between \ac{ASC} and \ac{HC}. The use of the \ac{ABIDE I} database, as observed in this study, does not yield to significant differences in structural brain patterns of heterogeneous disorders such as autism.

\section{Conclusions}
Throughout this study, our objective was to identify specific brain regions that could serve as reliable biomarkers in \ac{ASC}, utilizing the diverse and multisite \ac{ABIDE I} database. The images were analyzed by permutation tests and \ac{SAM} and \ac{SPM}, studying the influence of several factors, such as the presence of noisy images, determining differences by center or by condition. The results were not consistent, exhibiting those obtained with the \ac{SAM} method, when compared to the \ac{SPM} method, a much closer resemblance to the findings presented throughout the existing literature.

The main finding of the present analysis suggests that no significant differences can be observed in gray matter tissues of any specific brain region between individuals with \ac{ASC} and \ac{HC} subjects in structural MRI images of the \ac{ABIDE I} database. This finding does not definitively dismiss the possibility of the existence of specific structural differences implicated in autism. However, it does suggest that the structural MRI data provided by the \ac{ABIDE I} database, given its inherent heterogeneity, may not offer the best solution for investigating this phenomenon.

\section*{Acklowledgements}

This work was supported by the MCIN/ AEI/10.13039/501100011033/ and FEDER ``Una manera de hacer Europa''  under the PID2022-137451OB-I00  project, by the Consejer{\'i}a de Econom{\'i}a, Innovaci{\'o}n, Ciencia y Empleo (Junta de Andaluc{\'i}a) and FEDER under CV20-45250, A-TIC-080-UGR18, B-TIC-586-UGR20 and P20-00525 projects.

\appendix
\section{}

\subsection{Hypothesis tests}
The study involves conducting a hypothesis test, where the null hypothesis ($H_0$) represents the accepted belief thus far, which assumes no brain differences between individuals with autism and those without autism. The alternative hypothesis ($H_1$), the contrast to the null hypothesis, suggests the presence of significant differences in the brains of autistic and non-autistic individuals. 

The significance level $\alpha$ is set at the standard value of $0.05$, indicating the threshold for determining statistical significance. During the statistical analysis of the brain images, if the probability or $p-$value of the obtained results for the different brain regions is found to be less than the predetermined significance level $\alpha$ (\textless0.05), it would indicate that the observed findings are highly unlikely to occur by chance alone. In such a case, the null hypothesis would be rejected, and it would be concluded that there are statistically significant brain differences. 

\subsection{Selection of PLS dimension}
\label{pls_dimension}
\begin{figure}
    \centering
    \subfloat{
        \label{SAM all centers pls 1}
        \includegraphics[width=0.5\textwidth]{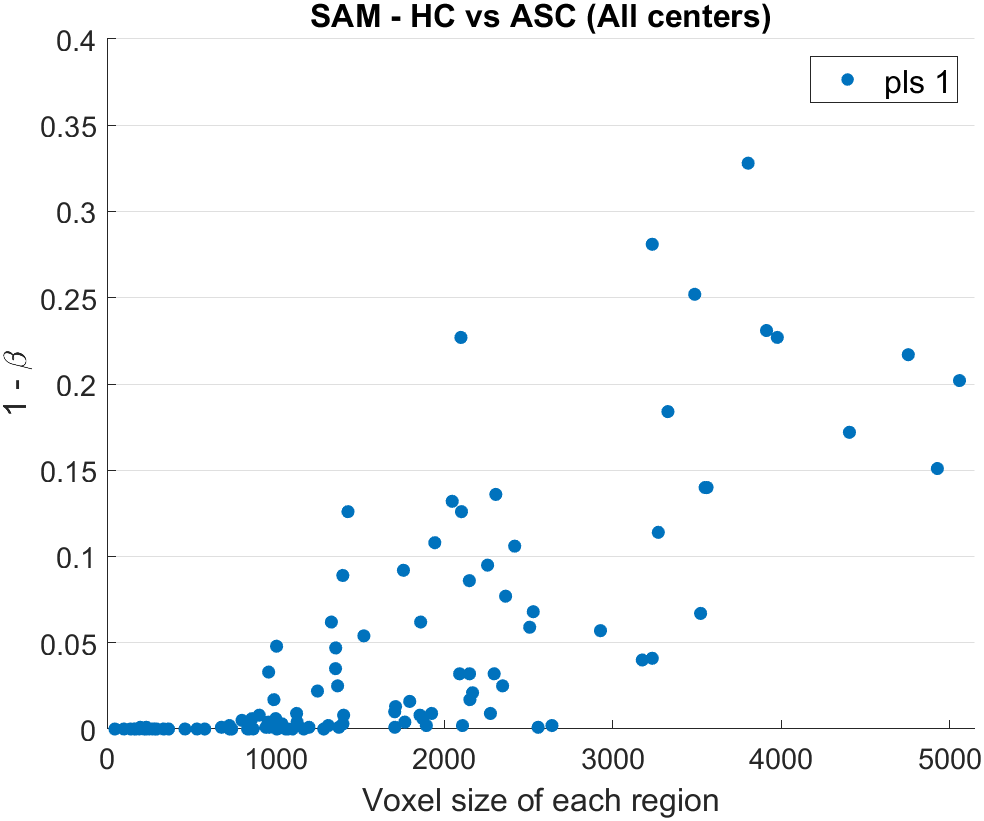}}   
    \subfloat{
        \label{SAM all centers pls 2}
        \includegraphics[width=0.5\textwidth]{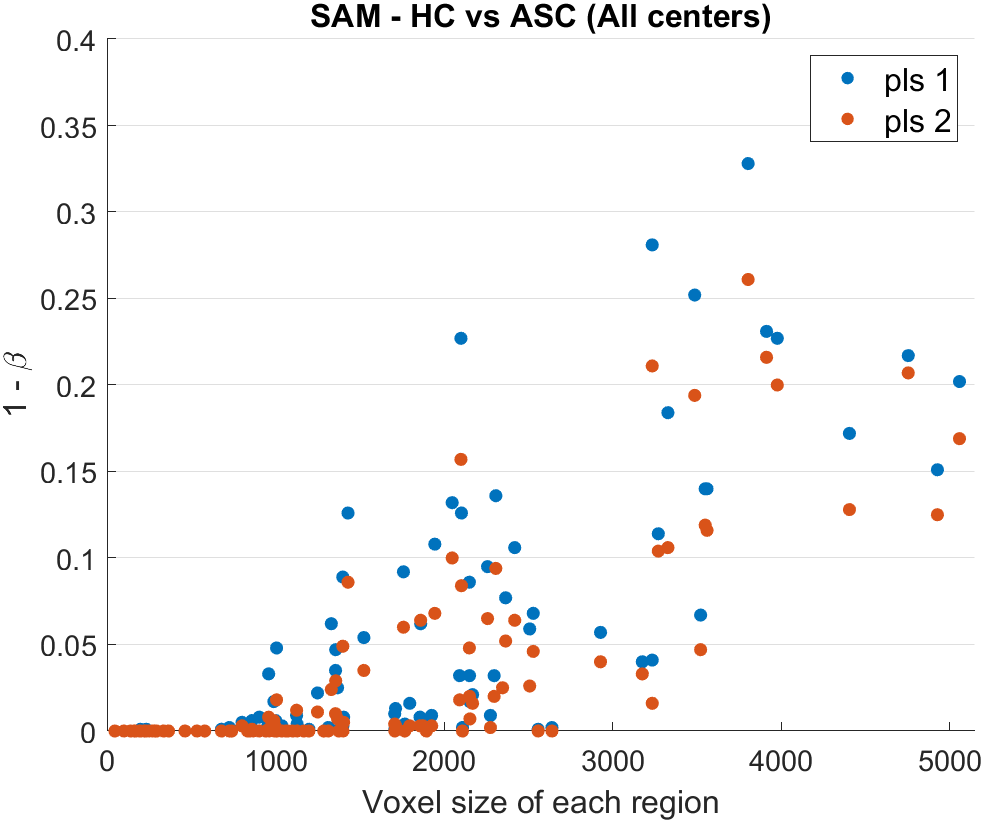}}

    \subfloat{
        \label{SAM all centers pls 5}
        \includegraphics[width=0.5\textwidth]{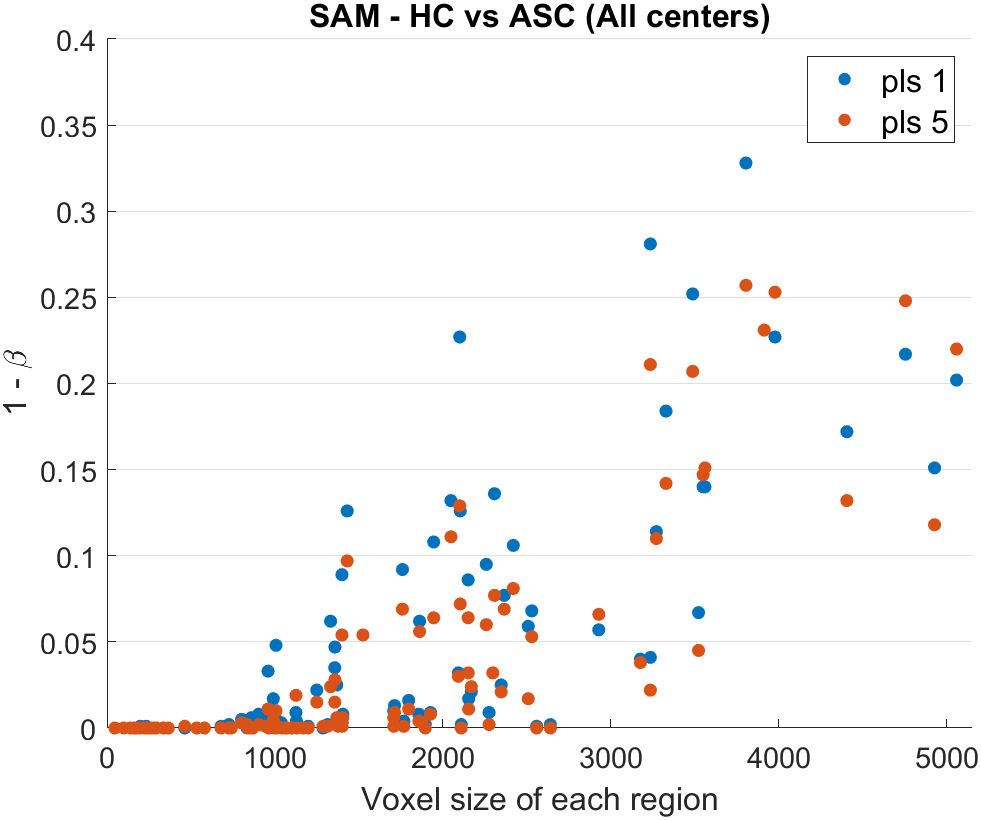}}   
    \subfloat{
        \label{SAM all centers pls 10}
        \includegraphics[width=0.5\textwidth]{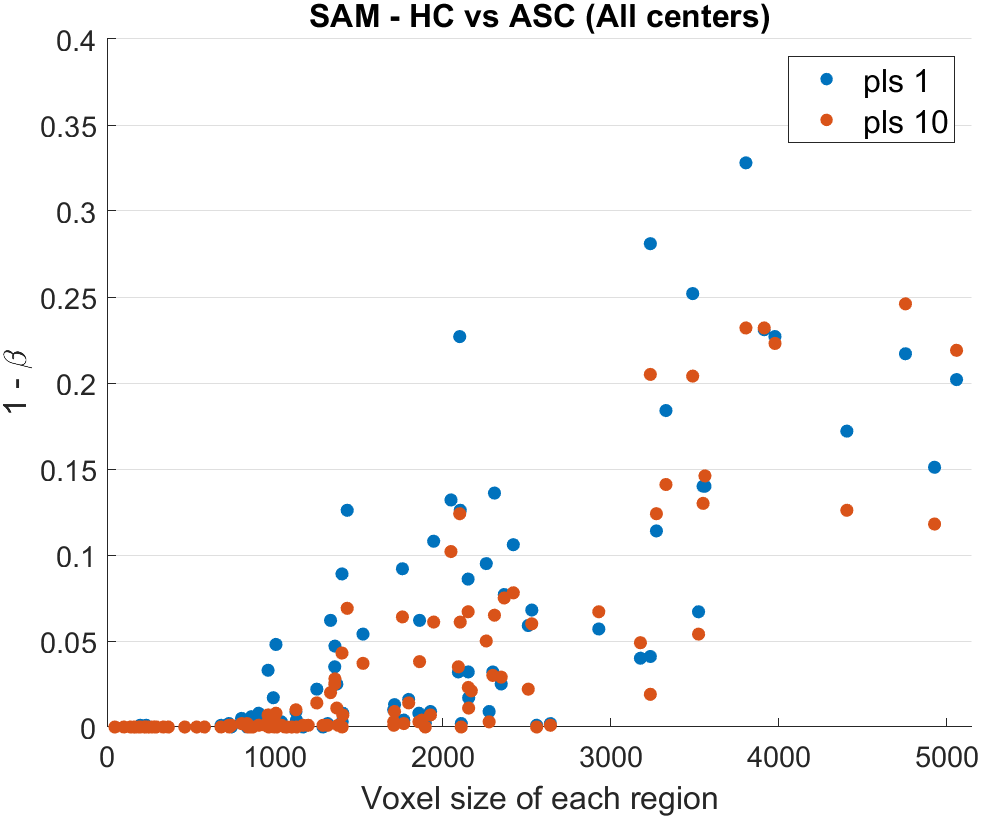}}
    \caption{Subplot on comparison between different values of dimensions for \ac{PLS} as a method of feature extraction in \ac{SAM} analysis for all centers \ac{HC} vs \ac{ASC} comparison.}\label{All centers pls}
\end{figure}

\begin{figure}
    \centering
    \subfloat{
        \label{SAM NYU center pls 1}
        \includegraphics[width=0.5\textwidth]{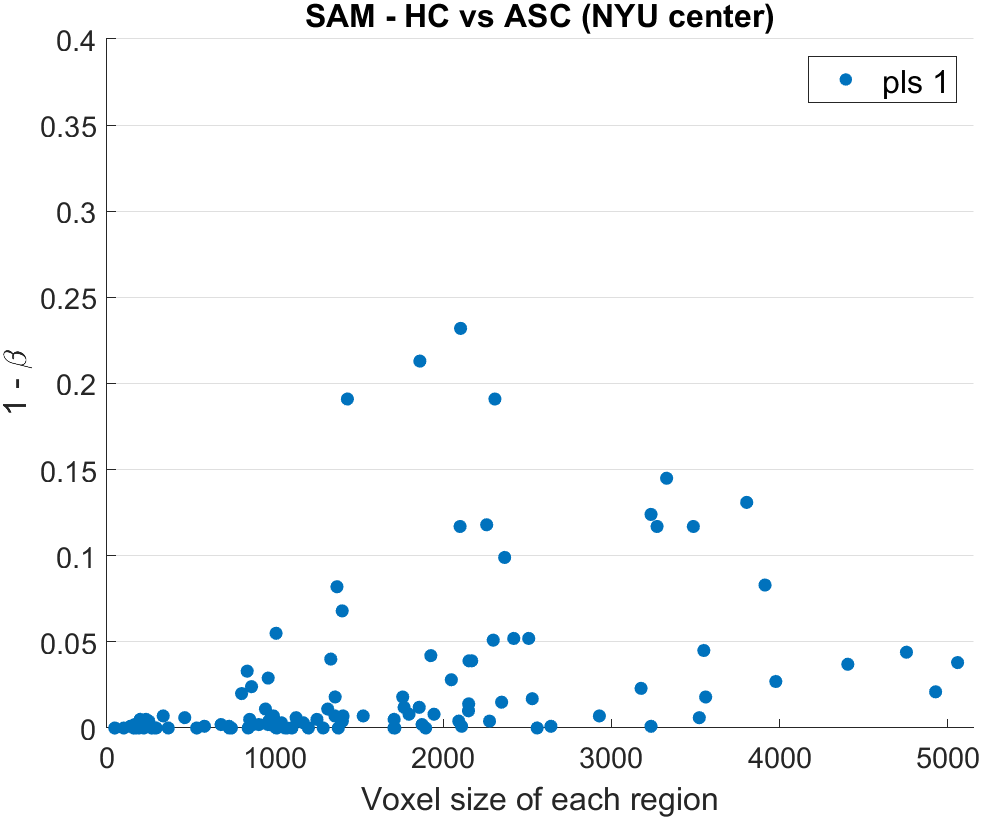}}   
    \subfloat{
        \label{SAM NYU centers pls 2}
        \includegraphics[width=0.5\textwidth]{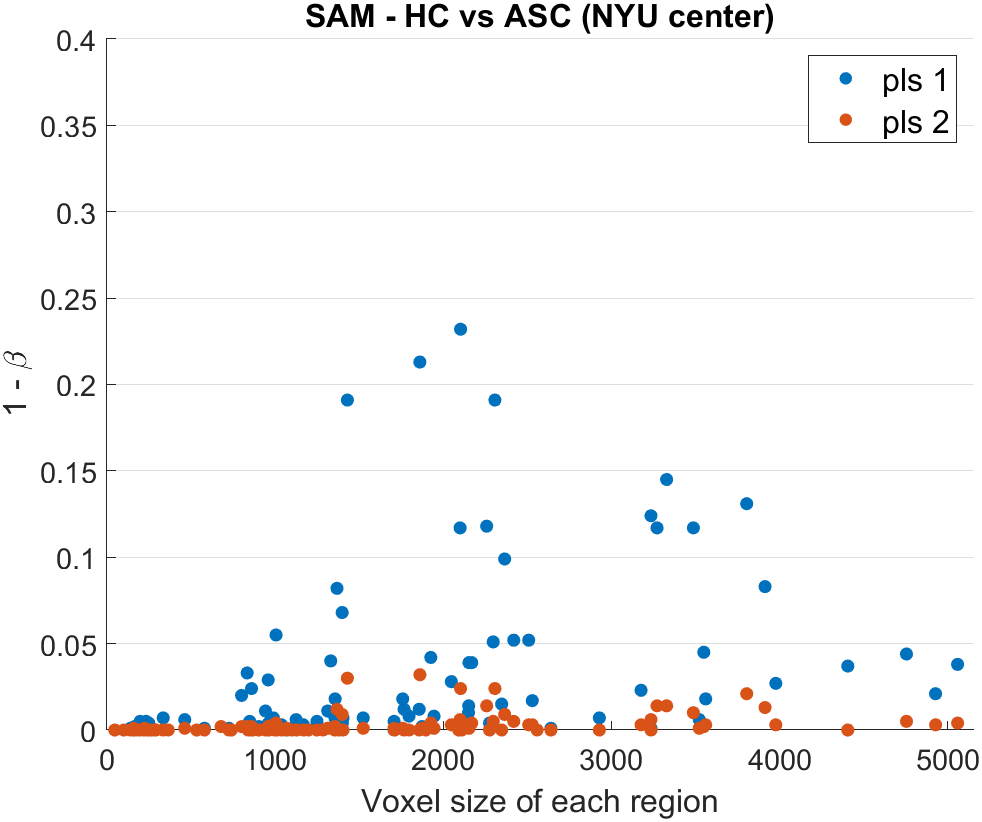}}

    \subfloat{
        \label{SAM all centers pls 3}
        \includegraphics[width=0.5\textwidth]{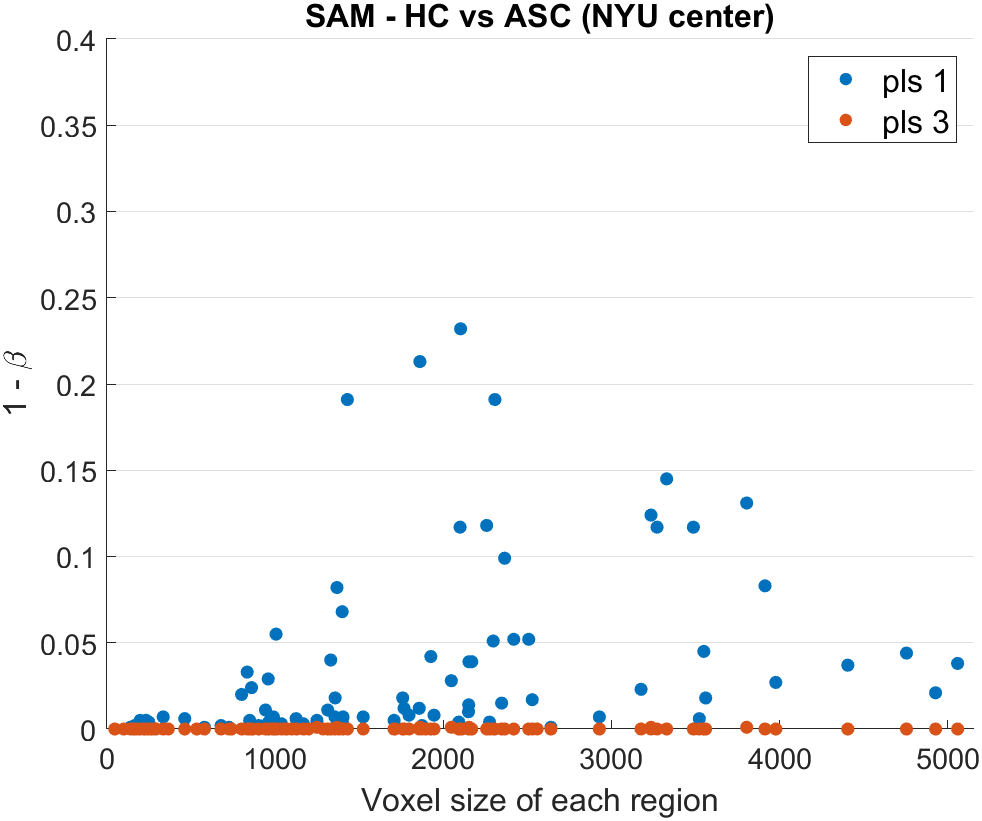}}
    \caption{Subplot on comparison between different values of dimensions for \ac{PLS} as a method of feature extraction in \ac{SAM} analysis for NYU center \ac{HC} vs \ac{ASC} comparison.}\label{NYU center pls} 
\end{figure}

\ac{SAM} employs a \ac{FES} stage on each \ac{ROI} within the images. In this study, the feature extraction method chosen is \ac{PLS}, with the dimension set to 1. The selection of 1 as the dimension of \ac{PLS} is based on the observation that increasing the number of coefficients or dimensions leads to less significant results, as depicted in Figures \ref{All centers pls} and \ref{NYU center pls}. These figures illustrate the \ac{HC} vs \ac{ASC} comparisons conducted on both the entire image dataset according to their condition (Figure \ref{All centers pls}) and the individual dataset from the NYU center (Figure \ref{NYU center pls}). Only this comparison is presented since it is of utmost importance in this study, and the NYU center is highlighted due to its substantial individual sample size.

As the dimension of \ac{PLS} increases, the frequency values of the brain regions generally decline, particularly in the case of the NYU center where this reduction is more pronounced. 
This implies that the utilization of a large number of \ac{PLS} components poses a potential risk of overfitting in heterogeneous environments. This risk arises from the fact that \ac{PLS} aims to maximize the fisher discriminant ratio (FDR) between the two groups analyzed. Consequently, as the number of \ac{PLS} components increases, the selected features become more complex and adapted to the specific characteristics of the data. Therefore, the selection of 1 as the \ac{PLS} dimension is the most suitable option for the analysis throughout this study.

        
\bibliographystyle{plainnat}
\bibliography{bibliography}

\end{document}